\documentclass[11pt,letter]{article} 

\usepackage{citesort,amsmath,upref,amssymb,color,bm}
\usepackage{epsfig,graphicx,rotating}

\def\baselinestretch{1}
\renewcommand{\baselinestretch}{1.5}\small\normalsize

\allowdisplaybreaks

\newcommand{\E}[1]{{\rm E} \left \{#1 \right \}}

\newcommand{\tr}[1]{{\rm tr} \left \{#1 \right \}}

\newcommand{\remark}[1]{}

\def\IM{\Im}
\def\RE{\Re}
\def\C{\mathbb{C}}
\def\Z{\mathbb{Z}}
\def\R{\mathbb{R}}

\def\bproof{{\em Proof. }}
\def\eproof{\hfill \fbox{} \newline}

\definecolor{grau}{gray}{0.6}

\newtheorem{theorem}{Theorem}[section]
\newtheorem{corollary}[theorem]{Corollary}
\newtheorem{lemma}[theorem]{Lemma}
\newtheorem{definition}[theorem]{Definition}
\newtheorem{proposition}[theorem]{Proposition}

\renewcommand{\jmath}{j}

\newcommand{\be}{\begin{equation}}
\newcommand{\ee}{\end{equation}}

\newcommand{\xxi}{\bm{\xi}}
\newcommand{\eeta}{\bm{\eta}}
\newcommand{\cchi}{\bm{\chi}}

\begin{document}

\title{\bf\vspace*{-11mm}  Complex-Valued Random Vectors and Channels:\\[-3.3mm]
          Entropy, Divergence, and Capacity\thanks{This
work was supported by WWTF grants MOHAWI (MA 44) and SPORTS (MA 07-004) as
well as by FWF grant ``Statistical Inference'' (S10603-N13) within the
National Research Network SISE. Parts of this work have been previously
published in \cite{Tauboeck02,Tauboeck04}.}\vspace*{-0mm}}

\author{\it Georg Taub\"{o}ck\\[2mm]
{\small Institute of Telecommunications, Vienna University of Technology}\\[-3.3mm]
{\small Gusshausstrasse\ 25/389, 1040 Vienna, Austria}\\[-3.3mm]
{\small Phone: +43 1 58801 38962, Fax: +43 1 58801 38999, E-mail:
gtauboec@nt.tuwien.ac.at}\vspace*{5mm} }

 \date{
 first revision; submitted to IEEE Trans. Inf. Theory, October 31, 2010}

\maketitle

\renewcommand{\baselinestretch}{1.22}\small\normalsize

\vspace*{-5mm}

\begin{quote}%
{\bf Abstract}---Recent research has demonstrated significant achievable
performance gains by exploiting circularity/non-circularity or
propeness/improperness of com\-plex-valued signals. In this paper, we
investigate the influence of these properties on important information
theoretic quantities such as entropy, divergence, and capacity. We prove two
maximum entropy theorems that strengthen previously known results. The proof
of the former theorem is based on the so-called \emph{circular analog} of a
given complex-valued random vector. Its introduction is supported by a
characterization theorem that employs a minimum Kullback-Leibler divergence
criterion. In the proof of latter theorem, on the other hand, results about
the second-order structure of complex-valued random vectors are exploited.
Furthermore, we address the capacity of multiple-input multiple-output (MIMO)
channels. Regardless of the specific distribution of the channel parameters
(noise vector and channel matrix, if modeled as random), we show that the
capacity-achieving input vector is circular for a broad range of MIMO
channels (including coherent and noncoherent scenarios). Finally, we
investigate the situation of an improper and Gaussian distributed noise
vector. We compute both capacity and capacity-achieving input vector and show
that improperness increases capacity, provided that the complementary
covariance matrix is exploited. Otherwise, a capacity loss occurs, for which
we derive
an explicit expression.\\[4mm]

\vspace*{10mm}
\end{quote}
\renewcommand{\baselinestretch}{1.5}\small\normalsize

\section{{Introduction}}\label{sec:intro}
Complex-valued signals are central in many scientific fields including
communications, array processing, acoustics and optics, oceanography and
geophysics, machine learning, and biomedicine. In recent research---for an
comprehensive overview see \cite{schreier_book}---it has been shown that
exploiting circularity/properness of complex-valued signals or lack of it
(non-circularity/improperness) is able to significantly enhance the
performance of the applied signal processing techniques. More specifically,
for the field of communications, it has been observed that important digital
modulation schemes including binary phase shift keying (BPSK), pulse
amplitude modulation (PAM), Gaussian minimum shift keying (GMSK), offset
quaternary phase shift keying (OQPSK), and \emph{baseband} (but not passband)
orthogonal frequency division multiplexing (OFDM), which is commonly called
discrete multitone (DMT), (potentially) produce non-circular/improper complex
baseband signals, see e.g.,
\cite{yoon97,gelli00,lampe02,gerst03,taub_comp_glob03,taub_comp_sp07}.
Non-circular/improper baseband communication signals can also arise due to
imbalance between their in-phase and quadrature (I/Q) components, and several
techniques for compensating for I/Q imbalance have been proposed
\cite{antt08,ryk08,zou08}.

Information theory, on the other hand, addresses fundamental performance
limits of communication systems and also has large impact on many other
scientific areas, where stochastic models are used. Therefore, results about
the most relevant information theoretic concepts, such as entropy,
divergence, and capacity, are of special interest. 
Clearly, information theory has the potential to study the performance limits
of signal processing algorithms and communications systems that exploit
circularity/properness or non-circularity/improperness. However, results in
this direction are limited and have to be investigated further. A significant
disadvantage of available results is that they often stick to a Gaussian
assumption, something which is non always the case in practice.

Apparently the first information theoretic result in this context analyzes
the differential entropy of complex-valued random vectors \cite{neeser93}.
More specifically, the maximum entropy theorem in \cite{neeser93} shows that
the differential entropy of a zero-mean complex-valued random vector with
given covariance matrix is upper bounded by the differential entropy of a
circular (and, consequently, zero mean and proper) Gaussian distributed
complex-valued random vector with the same covariance matrix. Using this
result, capacity results for vector-valued (multiple-input multiple-output;
MIMO) channels with complex-valued input and complex-valued output and
additive circular/proper Gaussian noise have been derived \cite{Telatar95}.
In particular, it has been shown that the capacity-achieving input vector is
Gaussian distributed and circular/proper.

Let us suppose that we are dealing with a complex-valued random vector which
is known to be non-Gaussian. In this situation, the upper bound on its
differential entropy given by the mentioned maximum entropy theorem turns out
to be not tight. The same is the case, if the complex-valued random vector is
known to be improper. Hence, there are two sources that decrease the
differential entropy of a complex-valued random vector, i.e., non-Gaussianity
and improperness. An important contribution of this paper is to derive
improved/tighter maximum entropy theorems for both situations.

The maximum entropy theorem in \cite{neeser93} associates a circular random
vector (i.e., the Gaussian distributed one) to a given complex-valued random
vector. As pointed out, this choice does not always lead to the the smallest
change in differential entropy. This raises the question, how we can
associate a circular random vector to an (in general) non-circular
complex-valued random vector in a canonical way but not forcing it to be
Gaussian distributed. The choice we propose is intuitive, and is furthermore
supported by a characterization theorem that is based on a minimum
Kullback-Leibler divergence criterion. It also leads to the desired improved
entropy upper bound for the case for which the random vector is known to be
non-Gaussian. A study of further properties complements its analysis.

%
%

As already mentioned, the maximum entropy theorem in \cite{neeser93} does not
yield a tight upper bound as well if the random vector is known to be
improper. Extending our work of \cite{Tauboeck02,Tauboeck04}, we derive an
improved maximum entropy theorem which addresses this situation. As a
by-product, we obtain a criterion for a matrix to be a valid complementary
covariance matrix (also termed pseudo-covariance matrix \cite{neeser93}). We
note that after our initial work \cite{Tauboeck02,Tauboeck04} the obtained
characterization of complementary covariance matrices has been extended in
\cite{Schreier05,schreier_spl06}. Meanwhile, expressions for the differential
entropy of improper Gaussian random vectors have appeared in literature as
well \cite{Koivunen_IT06,schreier_book}.

Finally, we apply the obtained improved maximum entropy theorems to derive
novel capacity results for complex-valued MIMO channels with additive noise
vectors. Without making use of any Gaussian assumption (in contrast to
\cite{Telatar95}), we show that capacity is achieved by circular random
vectors for a broad range of channels. These results include both the case of
a deterministic channel matrix and the case of a random channel matrix, which
is assumed to be either known to the receiver (coherent capacity) or unknown
(incoherent capacity). On the other hand, we investigate the capacity of
channels, whose noise is non-circular/improper and Gaussian distributed. Such
channels have been shown to occur if, e.g., DMT is used as modulation scheme
\cite{taub_comp_glob03,taub_comp_sp07}. Note that DMT is currently employed
in several xDSL standards \cite{xDSL_overview97}. We derive capacity
expressions for two cases: (i) we assume that the knowledge of the
complementary covariance matrix is taken into account (both at transmitter
and receiver); (ii) we assume that it is erroneously believed---i.e., that
the transceiver is designed assuming---that the noise has a vanishing
complementary covariance matrix, so that the information contained in the
complementary covariance matrix is ignored. This results in a decreased
capacity and we calculate the occurring capacity loss.

\noindent{\bf Notation.} The $n \times n$ identity matrix is denoted by
$\mathbf{I}_{n}$. We use the superscript $[\cdot]^{T}$ for transposition and
the superscript $[\cdot]^{H} \triangleq \left([\cdot]^{T}\right)^{*}$ for
Hermitian transposition, where the superscript $[\cdot]^{*}$ stands for
complex conjugation. $\jmath = \sqrt{-1}$ denotes the imaginary unit,  $\RE\{
\cdot\}$ and $\IM\{ \cdot\}$ are real and imaginary part, respectively, and
$\E{\cdot}$ refers to usual expectation. Throughout the paper, $\log(\cdot)$
denotes the logarithm taken with respect to an arbitrary but fixed base.
Therefore, all results are valid regardless of the chosen unit for
differential entropy (\emph{nats} or \emph{bits}).

\noindent{\bf Outline.} The remainder of this paper is organized as follows.
In Section \ref{sec:preliminaries}, we introduce our framework and present
initial results about the distribution and second-order properties of
complex-valued random vectors. Section \ref{sec:analogs} deals with the
question, how to circularize complex-valued random vectors and analyzes the
proposed method. The differential entropy of complex-valued random vectors is
addressed in Section \ref{sec:entropy} and two improved maximum entropy
theorems are proved. Finally, in Section \ref{sec:capacity}, we present
various capacity results for complex-valued MIMO channels.

\section{Framework and Preliminary Results}\label{sec:preliminaries}

We consider complex-valued random vectors $\mathbf{x} \in \C^{n}$. We assume
that $\mathbf{x}^{\text{(r)}} \in \R^{2n}$, where $\mathbf{x}^{\text{(r)}}
\triangleq \left[\RE\{\mathbf{x}^{T}\}\,\, \IM\{\mathbf{x}^{T}\}\right]^{T}$
is defined by stacking of real and imaginary part of $\mathbf{x}$, is
distributed according to a joint multivariate $2n$-dimensional probability
density function (pdf) $f_{\mathbf{x}^{\text{(r)}}}(\xxi)$. More precisely,
it is assumed that the measure\footnote{Here, we refer to the measure defined
on the Borel $\sigma$-field on $\R^{2n}$ induced by the measurable function
defining the random vector.} defining the distribution of
$\mathbf{x}^{\text{(r)}}$ is absolutely continuous with respect to
$\lambda_{2n}$, where $\lambda_{2n}$ denotes the $2n$-dimensional Lebesgue
measure \cite{Hal74}. Accordingly, whenever an integral appears in this
paper, integration is meant with respect to the Lebesgue measure of
appropriate dimension. Note that when we refer to the distribution of
$\mathbf{x}$, we mean the distribution of $\mathbf{x}^{\text{(r)}}$ defined
by the pdf $f_{\mathbf{x}^{\text{(r)}}}(\xxi)$. Hence, a complex-valued
random vector $\mathbf{x}$ will be called Gaussian distributed if
$\mathbf{x}^{\text{(r)}}$ is (multivariate) Gaussian distributed.
\begin{definition}
A complex-valued random vector \emph{$\mathbf{x} \in \C^{n}$} is said to be
\emph{circular}, if \emph{$\mathbf{x}$} has the same distribution as
\emph{$e^{\jmath 2 \pi \theta}\mathbf{x}$} for all \emph{$\theta \in
[0,1[$}\,, otherwise it is said to be \emph{non-circular}. The set of all
circular complex-valued random vectors \emph{$\mathbf{x} \in \C^{n}$}, whose
distribution is absolutely continuous with respect to \emph{$\lambda_{2n}$},
is denoted by \emph{$\mathcal{C}_{n}$}.
\end{definition}
It is well known, see e.g.,
\cite{neeser93,taub_diss05,schreier_book,Tauboeck02}, that for a complete
second-order characterization of a complex-valued random vector $\mathbf{x}
\in \C^{n}$ not only mean vector $\mathbf{m}_{\mathbf{x}}\triangleq
\E{\mathbf{x}}$ and covariance matrix $\mathbf{C}_{\mathbf{x}}\triangleq
\E{(\mathbf{x}-\right.$ $\left.\mathbf{m}_{\mathbf{x}})
(\mathbf{x}-\mathbf{m}_{\mathbf{x}})^{H}}$ but also \emph{complementary
covariance matrix} $\mathbf{P}_{\mathbf{x}}\triangleq
\E{(\mathbf{x}-\mathbf{m}_{\mathbf{x}})
(\mathbf{x}-\mathbf{m}_{\mathbf{x}})^{T}}$ are required. 
Note that
both mean vector and complementary covariance matrix of a circular
complex-valued random vector are vanishing provided that its first- and
second-order moments exist \cite{schreier_book}.
\begin{definition}
A complex-valued random vector $\mathbf{x} \in \C^{n}$ is said to be
\emph{proper}, if its complementary covariance matrix vanishes, otherwise it
is said to be \emph{improper}.
\end{definition}
Hence, circularity implies properness (under the assumption of existing
first- and second-order moments). 
Note that a zero-mean and proper Gaussian random vector is circular.

\subsection{Polar and Sheared-Polar Representation}

Here, we present some auxiliary results about the distribution of
complex-valued random vectors. Let us denote by
$\mathbb{T}^{\text{(p$\rightarrow$\,r)}}$ the mapping
\[
    \mathbb{T}^{\text{(p$\rightarrow$\,r)}}:
    \left\{ \begin{array}{lcl}
    (\R_{0}^{+})^{n}\times ([0,1[)^{n} & \rightarrow & \R^{2n}, \\
    {[r_1 \cdots r_n\, \, \phi_1 \cdots \phi_n]^T} & \mapsto & \left[r_1 \cos (2 \pi \phi_1)
    \cdots r_n \cos (2 \pi \phi_n) \, \, r_1 \sin(2 \pi \phi_1)  \cdots r_n \sin (2 \pi \phi_n)\right]^T,
    \end{array}\right.
\]
where $\R_{0}^{+}$ denotes the set of non-negative reals. There exists the
inverse $\mathbb{T}^{\text{(r$\rightarrow$\,p)}} \triangleq
\left({\mathbb{T}^{\text{(p$\rightarrow$\,r)}}}\right)^{-1}$, provided that
we set $\phi_i\triangleq 0$ for $r_i=0$, $i=1,\ldots,n$. Note that the set,
by which the domain of $\mathbb{T}^{\text{(p$\rightarrow$\,r)}}$ is reduced
according to this convention has measure zero with respect to $\lambda_{2n}$.
In the following, $\mathbf{x}^{\text{(r)}}$ will be called \emph{real
representation} of $\mathbf{x}$, whereas $\mathbf{x}^{\text{(p)}}\triangleq
\mathbb{T}^{\text{(r$\rightarrow$\,p)}}(\mathbf{x}^{\text{(r)}})$ will be
denoted as \emph{polar representation} of $\mathbf{x}$.
\begin{lemma}\label{lem:polar_distr}
Suppose $\mathbf{x} \in \C^{n}$ is a complex-valued random vector, which is
distributed according to the pdf \emph{$f_{\mathbf{x}^{\text{(r)}}}(\xxi)$}.
Then, the pdf of its polar representation \emph{$\mathbf{x}^{\text{(p)}}$} is
given by\emph{
\begin{align*}
    f_{\mathbf{x}^{\text{(p)}}}(r_1,\ldots,r_n,\phi_1,\ldots,\phi_n)&= \left\{
    \begin{array}{ll}
      (2\pi)^{n} (r_1 \cdots r_n) f_{\mathbf{x}^{\text{(r)}}}\left(\mathbb{T}^{\text{(p$\rightarrow$\,r)}}(r_1,\ldots,r_n,\phi_1,\ldots,\phi_n)\right), & (r_1,\ldots\\
       &\hspace*{-35mm} \ldots,r_n,\phi_1,\ldots,\phi_n) \in (\R_{0}^{+})^{n}\times ([0,1[)^{n}\\
       0, & \text{otherwise}
    \end{array} \right.
\end{align*}}
almost everywhere with respect to \emph{$\lambda_{2n}$}
(\emph{$\lambda_{2n}\text{-a.e.}$}) \emph{\cite{Hal74}}.
\end{lemma}
\bproof Follows from
\begin{align*}
\int\limits_{\mathcal{A}} \hspace*{0mm} f_{\mathbf{x}^{\text{(p)}}}(\xxi)d\xxi \hspace*{3mm} = \hspace*{-6mm} \int\limits_{\hspace*{7mm}\mathbb{T}^{\text{(p$\rightarrow$\,r)}}(\mathcal{A})}
\hspace*{-9mm} f_{\mathbf{x}^{\text{(r)}}}(\xxi)d\xxi \hspace*{3mm} = \hspace*{3mm} \int\limits_{\mathcal{A}} \hspace*{0mm} f_{\mathbf{x}^{\text{(r)}}}\left(\mathbb{T}^{\text{(p$\rightarrow$\,r)}}(\xxi)\right)
\left|J_{\mathbb{T}^{\text{(p$\rightarrow$\,r)}}}(\xxi) \right|d\xxi,
\end{align*}
for all Lebesgue measurable sets $\mathcal{A} \subset
(\R_{0}^{+})^{n}\times([0,1[)^{n}$, where the Jacobian determinant
$J_{\mathbb{T}^{\text{(p$\rightarrow$\,r)}}}(\xxi)$ of
$\mathbb{T}^{\text{(p$\rightarrow$\,r)}}$ is easily computed as
$J_{\mathbb{T}^{\text{(p$\rightarrow$\,r)}}}(r_1,\ldots,r_n,\phi_1,\ldots,\phi_n)=(2\pi)^{n}
(r_1 \cdots r_n)$ \cite{Rud87}.
 \eproof
 Observing that the pdf of $\mathbf{y}^{\text{(p)}}_{(\theta)}$ of the random vector $\mathbf{y}_{(\theta)} \triangleq e^{\jmath 2
\pi \theta}\mathbf{x}$ (with $\theta\in [0,1[$ being deterministic) satisfies
$f_{\mathbf{y}^{\text{(p)}}_{(\theta)}}(r_1,\ldots,r_n,\phi_1,\ldots,\phi_n)
=f_{\mathbf{x}^{\text{(p)}}}\big(r_1,\ldots,r_n,[\phi_1-\theta]_{[0,1[},\ldots,[\phi_n-\theta]_{[0,1[}\big)$\hspace*{1mm}$\lambda_{2n}\text{-a.e.}$,
where the notation $[\cdot]_{[0,1[}$ is shorthand for modulo with respect to
the interval $[0,1[$, we obtain the following corollary.
\begin{corollary}\label{cor:circ_polar}
A complex-valued random vector \emph{$\mathbf{x} \in \C^{n}$} is circular if
and only if the pdf of its \emph{polar representation}
\emph{$\mathbf{x}^{\text{(p)}}$} satisfies \emph{
\begin{align*}
    f_{\mathbf{x}^{\text{(p)}}}(r_1,\ldots,r_n,\phi_1,\ldots,\phi_n)
=f_{\mathbf{x}^{\text{(p)}}}\big(r_1,\ldots,r_n,[\phi_1-\theta]_{[0,1[},\ldots,[\phi_n-\theta]_{[0,1[}\big) \quad \forall \, \theta \in [0,1[ \quad \lambda_{2n}\text{-a.e.}.\end{align*}}
\end{corollary}
Let us denote by $\mathbb{T}^{\text{(s$\rightarrow$\,p)}}$ the mapping
\[
    \mathbb{T}^{\text{(s$\rightarrow$\,p)}}:
    \left\{ \begin{array}{lcl}
    (\R_{0}^{+})^{n}\times ([0,1[)^{n} & \rightarrow & (\R_{0}^{+})^{n}\times ([0,1[)^{n}, \\
    {[r_1 \cdots r_n\, \, \phi_1 \cdots \phi_n]^T} & \mapsto & \left[r_1 \cdots r_n\,   [\phi_1+\phi_n]_{[0,1[} \cdots [\phi_{n-1}+\phi_n]_{[0,1[}\,\, \phi_n\right]^T,
    \end{array}\right.
\]
which is one-to-one with inverse $\mathbb{T}^{\text{(p$\rightarrow$\,s)}}
\triangleq \left({\mathbb{T}^{\text{(s$\rightarrow$\,p)}}}\right)^{-1}$ given
by
\[
    \mathbb{T}^{\text{(p$\rightarrow$\,s)}}:
    \left\{ \begin{array}{lcl}
    (\R_{0}^{+})^{n}\times ([0,1[)^{n} & \rightarrow & (\R_{0}^{+})^{n}\times ([0,1[)^{n}, \\
    {[r_1 \cdots r_n\, \, \phi_1 \cdots \phi_n]^T} & \mapsto & \left[r_1 \cdots r_n\,   [\phi_1-\phi_n]_{[0,1[} \cdots [\phi_{n-1}-\phi_n]_{[0,1[}\,\, \phi_n\right]^T.
    \end{array}\right.
\]
This follows immediately from the identity
\begin{equation}\label{mod_identity}
    [\phi]_{[0,1[}=\phi+n(\phi), \quad \phi \in \R,
\end{equation}
where $n(\phi) \in \Z$. In the following, $\mathbf{x}^{\text{(s)}}\triangleq
\mathbb{T}^{\text{(p$\rightarrow$\,s)}}(\mathbf{x}^{\text{(p)}}) =
\mathbb{T}^{\text{(p$\rightarrow$\,s)}}\left(\mathbb{T}^{\text{(r$\rightarrow$\,p)}}(\mathbf{x}^{\text{(r)}})\right)$
will be called \emph{sheared-polar representation} of $\mathbf{x}$.
\begin{lemma}\label{lem:sheared_distr}
Suppose $\mathbf{x} \in \C^{n}$ is a complex-valued random vector. Then, the
pdfs of its polar representation \emph{$\mathbf{x}^{\text{(p)}}$} and its
sheared-polar representation \emph{$\mathbf{x}^{\text{(s)}}$} are related
according to \emph{
\begin{align*}
    f_{\mathbf{x}^{\text{(s)}}}(r_1,\ldots,r_n,\phi_1,\ldots,\phi_n) &= f_{\mathbf{x}^{\text{(p)}}}\left(r_1,\ldots,r_n, [\phi_1+\phi_n]_{[0,1[},\ldots,[\phi_{n-1}+\phi_n]_{[0,1[},\phi_n\right) \quad \lambda_{2n}\text{-a.e.},\\
    f_{\mathbf{x}^{\text{(p)}}}(r_1,\ldots,r_n,\phi_1,\ldots,\phi_n) &= f_{\mathbf{x}^{\text{(s)}}}\left(r_1,\ldots,r_n,[\phi_1-\phi_n]_{[0,1[},\ldots,[\phi_{n-1}-\phi_n]_{[0,1[},\phi_n\right) \quad \lambda_{2n}\text{-a.e.}.
\end{align*}}
\end{lemma}
\bproof Observe that the measure defining the distribution of
$\mathbf{x}^{\text{(s)}}$ is absolutely continuous with respect to
$\lambda_{2n}$, since $\lambda_{2n}(\mathcal{N})=0$ implies
$\lambda_{2n}(\mathbb{T}^{\text{(s$\rightarrow$\,p)}}(\mathcal{N}))=0$ for
all $\mathcal{N} \subset (\R_{0}^{+})^{n}\times ([0,1[)^{n}$, as can be seen
by distinction of cases according to the modulo-$[0,1[$ operation. Since
$\mathbb{T}^{\text{(s$\rightarrow$\,p)}}$ is not continuous on its whole
domain, we define the auxiliary mapping
$\widetilde{\mathbb{T}}^{\text{(s$\rightarrow$\,p})}\!:$
$(\R_{0}^{+})^{n}\times \R^{n} \rightarrow (\R_{0}^{+})^{n}\times \R^{n},$
$[r_1 \cdots r_n\, \, \phi_1 \cdots \phi_n]^T \mapsto  \left[r_1 \cdots r_n\,
\, (\phi_1+\phi_n) \cdots (\phi_{n-1}\!+\phi_n)\,\, \phi_n \right]^T$, which
is one-to-one with inverse
$\widetilde{\mathbb{T}}^{\text{(p$\rightarrow$\,s})} \triangleq
\left(\widetilde{\mathbb{T}}^{\text{(s$\rightarrow$\,p})}\right)^{-1}$, where
$\widetilde{\mathbb{T}}^{\text{(p$\rightarrow$\,s})}\!:$
$(\R_{0}^{+})^{n}\times \R^{n} \rightarrow (\R_{0}^{+})^{n}\times \R^{n},$
$[r_1 \cdots r_n\, \, \phi_1 \cdots \phi_n]^T \mapsto  \left[r_1 \cdots r_n\,
\, (\phi_1-\phi_n) \cdots (\phi_{n-1}\!-\phi_n)\,\, \phi_n \right]^T$. Its
Jacobian determinant is identically
$J_{\widetilde{\mathbb{T}}^{\text{(s$\rightarrow$\,p})}}(r_1,\ldots,r_n,\phi_1,\ldots,\phi_n)\equiv
1$. Suppose $\mathcal{A} \subset (\R_{0}^{+})^{n}\times([0,1[)^{n}$ is any
set of the form $\mathcal{A} = [a_1,b_1[\times \cdots \times
[a_{2n},b_{2n}[$. From \eqref{mod_identity} it follows that there exists a
finite partition $\{\mathcal{A}_1,\ldots,\mathcal{A}_N\}$ of $\mathcal{A}$,
i.e., $\mathcal{A}=\bigcup\limits_{i=1}^{N} \mathcal{A}_i$ and $\mathcal{A}_i
\cap \mathcal{A}_j = \emptyset$ for $i\neq j$, such that
$\widetilde{\mathbb{T}}^{\text{(p$\rightarrow$\,s})}\left(\mathbb{T}^{\text{(s$\rightarrow$\,p)}}(\mathcal{A})\right)
= \bigcup\limits_{i=1}^{N} (\mathcal{A}_i+\mathbf{k}_i)$, where $\mathbf{k}_i
\in \{0\}^{n} \times \Z^{n-1} \times \{0\}$. Here,
$\mathcal{A}_i+\mathbf{k}_i$ denote the disjoint sets
$\mathcal{A}_i+\mathbf{k}_i \triangleq \{\xxi \in \R^{2n}: \xxi -
\mathbf{k}_i \in \mathcal{A}_i \}$, $i=1,\ldots,N$. Note that the partition
is caused by the modulo-$[0,1[$ operation used in the definition of
$\mathbb{T}^{\text{(s$\rightarrow$\,p)}}$ and corresponds to the required
distinction of cases when  investigating
$\widetilde{\mathbb{T}}^{\text{(p$\rightarrow$\,s})}\left(\mathbb{T}^{\text{(s$\rightarrow$\,p)}}(\mathcal{A})\right)$.
Therefore,
\begin{align*}
\int\limits_{\mathcal{A}} \hspace*{0mm} f_{\mathbf{x}^{\text{(s)}}}(\xxi)d\xxi \hspace*{3mm} &= \hspace*{-6mm} \int\limits_{\hspace*{7mm}\mathbb{T}^{\text{(s$\rightarrow$\,p)}}(\mathcal{A})}
\hspace*{-9mm} f_{\mathbf{x}^{\text{(p)}}}(\xxi)d\xxi \hspace*{3mm} = \hspace*{-16mm} \int\limits_{\hspace*{15mm} \widetilde{\mathbb{T}}^{\text{(p$\rightarrow$\,s})}\left(\mathbb{T}^{\text{(s$\rightarrow$\,p)}}(\mathcal{A})\right)} \hspace*{-19mm} f_{\mathbf{x}^{\text{(p)}}}\left(\widetilde{\mathbb{T}}^{\text{(s$\rightarrow$\,p})}(\xxi)\right)
\left|J_{\widetilde{\mathbb{T}}^{\text{(s$\rightarrow$\,p)}}}(\xxi) \right|d\xxi\\
&=\hspace*{2mm}\bigcup\limits_{i=1}^{N}\hspace*{-1mm}\int\limits_{\hspace*{2mm} \mathcal{A}_i+\mathbf{k}_i} \hspace*{-4mm} f_{\mathbf{x}^{\text{(p)}}}\left(\widetilde{\mathbb{T}}^{\text{(s$\rightarrow$\,p})}(\xxi)\right)
d\xxi \hspace*{3mm}= \hspace*{2mm}\bigcup\limits_{i=1}^{N}\hspace*{1mm}\int\limits_{\mathcal{A}_i} \hspace*{0mm} f_{\mathbf{x}^{\text{(p)}}}\left(\widetilde{\mathbb{T}}^{\text{(s$\rightarrow$\,p})}(\xxi+\mathbf{k}_i)\right)
d\xxi \\
&\overset{(*)}{=} \hspace*{2mm} \bigcup\limits_{i=1}^{N}\hspace*{1mm}\int\limits_{\mathcal{A}_i} \hspace*{0mm} f_{\mathbf{x}^{\text{(p)}}}\left(\mathbb{T}^{\text{(s$\rightarrow$\,p)}}(\xxi)\right)d\xxi
\hspace*{3mm}= \hspace*{2mm} \int\limits_{\mathcal{A}} \hspace*{0mm} f_{\mathbf{x}^{\text{(p)}}}\left(\mathbb{T}^{\text{(s$\rightarrow$\,p)}}(\xxi)\right)d\xxi,
\end{align*}
where $(*)$ follows from \eqref{mod_identity} and the fact that
$\widetilde{\mathbb{T}}^{\text{(s$\rightarrow$\,p})}(\xxi+\mathbf{k}_i) \in
(\R_{0}^{+})^{n}\times ([0,1[)^{n}$ for $\xxi \in \mathcal{A}_i$. This
implies the statement, see e.g., \cite{Hal74}. \eproof Combining Corollary
\ref{cor:circ_polar} and Lemma \ref{lem:sheared_distr}, while applying
\eqref{mod_identity}, yields the following corollary.
\begin{corollary}\label{cor:circ_sheared}
A complex-valued random vector \emph{$\mathbf{x} \in \C^{n}$} is circular if
and only if the pdf of its \emph{sheared-polar representation}
\emph{$\mathbf{x}^{\text{(s)}}$} does not depend on \emph{$\phi_n$}, i.e.,
\emph{
\begin{align*}
    f_{\mathbf{x}^{\text{(s)}}}(r_1,\ldots,r_n,\phi_1,\ldots,\phi_n)
=f_{\mathbf{x}^{\text{(s)}}}(r_1,\ldots,r_n,\phi_1,\ldots,\phi_{n-1}) \quad \lambda_{2n}\text{-a.e.}.\end{align*}}
\end{corollary}

\subsection{Second-Order Properties}
In the following, we establish some results about covariance and
complementary covariance matrices of complex-valued random vectors. For a
given complex-valued matrix $\mathbf{A}\in \C^{n \times m}$, let us denote by
$\overline{\mathbf{A}} \in \R^{2n \times 2m}$ and $\underline{\mathbf{A}}\in
\R^{2n \times 2m}$ the real-valued matrices
\begin{align}\label{matr_real_repr}
    \overline{\mathbf{A}} \triangleq \left[\begin{array}{lr}  \RE \{\mathbf{A} \} & -\IM \{\mathbf{A }\}\\
                      \IM \{\mathbf{A} \} & \RE \{\mathbf{A }\}  \end{array} \right] \quad \text{and} \quad
                      \underline{\mathbf{A}}\triangleq \left[\begin{array}{lr}  \RE \{\mathbf{A} \} & \IM \{\mathbf{A }\}\\
                      \IM \{\mathbf{A} \} & -\RE \{\mathbf{A }\}  \end{array} \right].
\end{align}
This notation allows a simple expression of the covariance matrix of the real
representation $\mathbf{C}_{\mathbf{x}^{\text{(r)}}}$ of a complex-valued
random vector $\mathbf{x}$ in terms of covariance matrix
$\mathbf{C}_{\mathbf{x}}$ and complementary covariance matrix
$\mathbf{P}_{\mathbf{x}}$ as
\cite{neeser93,taub_diss05,schreier_book,Tauboeck02}
\begin{align}\label{expr_cov_compl}
\mathbf{C}_{\mathbf{x}^{\text{(r)}}} = \frac{1}{2}
\overline{\mathbf{C}}_{\mathbf{x}} + \frac{1}{2}
\underline{\mathbf{P}}_{\mathbf{x}}.
\end{align}
Furthermore, $\mathbf{A}$, $\overline{\mathbf{A}}$, and
$\underline{\mathbf{A}}$ satisfy remarkable algebraic properties, as stated
by the next lemma.
\begin{lemma}\label{lem:algebraic_props}\emph{
    \begin{subequations}
   \begin{eqnarray}\label{alg_a}
        &&\hspace*{-38mm} \mathbf{C=AB} \hspace*{3mm} \Leftrightarrow \hspace*{3mm} \mathbf{\overline{C}=\overline{A}\,\overline{B}} \hspace*{3mm} \Leftrightarrow \hspace*{3mm}
        \mathbf{\underline{C}=\overline{A}\underline{B}} \\\label{alg_b}
        \mathbf{C=AB^{*}} & \Leftrightarrow & \mathbf{\underline{C}=\underline{A}{\overline{B}}}\\\label{alg_c}
        \mathbf{C}=\mathbf{A}^{H} & \Leftrightarrow & \mathbf{\overline{C}}=\mathbf{\overline{A}}^{T} \\\label{alg_d}
        \mathbf{U} \in {\C}^{\hspace{0.3mm} n \times n} \text{ unitary} & \Leftrightarrow & \mathbf{\overline{U}} \in {\R}^{\hspace{0.3mm} 2 n \times 2 n} \text{ orthonormal}\\\label{alg_e}
        &&\hspace*{-37mm}\det \mathbf{\overline{A}}  = \left | \det \mathbf{A}  \right | ^{2} = \det\left(\mathbf{A} \mathbf{A}^{H}\right), \quad \mathbf{A} \in {\C}^{\hspace{0.3mm} n \times n}
   \end{eqnarray}
\end{subequations}}
\end{lemma}
\bproof For some of the statements, see also \cite{Telatar95}. Direct
calculations yield \eqref{alg_a} and \eqref{alg_b}. \eqref{alg_c} follows
from the definition of $\overline{\mathbf{A}}$. A combination of
\eqref{alg_a} and \eqref{alg_c}, while observing that
$\overline{\mathbf{I}}_{n}=\mathbf{I}_{2n}$, yields \eqref{alg_d}. Finally,
for \eqref{alg_e},
\[
    \det \mathbf{\overline{A}}  = \det \left(\left[
\begin{array}{cr}
  \mathbf{I}_{n} & \jmath \mathbf{I}_{n} \\
  \mathbf{0} & \mathbf{I}_{n} \\
\end{array}
\right]\mathbf{\overline{A}} \left[
\begin{array}{cr}
  \mathbf{I}_{n} & - \jmath \mathbf{I}_{n} \\
  \mathbf{0} & \mathbf{I}_{n} \\
\end{array}
\right]\right)=\det\left[
\begin{array}{cc}
  \mathbf{A} & \mathbf{0} \\
  \IM \{\mathbf{A} \} & \mathbf{A^{*}} \\
\end{array}
\right]  = \det \mathbf{A} \det \mathbf{A}^{*}.
\]
\eproof We are especially interested in the eigenvalues of
$\underline{\mathbf{P}}_{\mathbf{x}}$. We will show that they are essentially
given by the \emph{singular values} of $\mathbf{P}_{\mathbf{x}}$. Note that
the \emph{singular value decomposition} (SVD) \cite{golub96} of a matrix
$\mathbf{A} \in \C^{n \times m}$ factorizes $\mathbf{A}$ into three matrices,
i.e., $\mathbf{A} = \mathbf{U} \mathbf{\Lambda} \mathbf{V}^{H}$. It is well
defined for all rectangular complex matrices and yields unitary matrices
$\mathbf{U} \in {\C}^{n \times n}$ and $\mathbf{V} \in {\C}^{m \times m}$ and
a diagonal matrix $\mathbf{\Lambda} \in {\R}^{n \times m}$, i.e.,
$\mathbf{\Lambda} = \mbox{diag}^{n \times
m}\left\{\lambda_{1},\ldots,\lambda_{\min\{n,m\}}\right\}$, with non-negative
entries on its main diagonal---the singular values. $\mathbf{U}$ and
$\mathbf{V}$ can be chosen such that the singular values are ordered in
descending order. In case the matrix $\mathbf{A} \in \C^{n \times n}$ is
symmetric (not Hermitian), i.e., $\mathbf{A}^{T}=\mathbf{A}$, there is a
special SVD known as \emph{Takagi factorization} \cite{Horn90}. It is given
by the factorization
\begin{align}\label{takagi_equation}
        \mathbf{A}=\mathbf{Q} \mathbf{\Lambda} \mathbf{Q}^{T},
\end{align}
where the columns of $\mathbf{Q}$ are the orthonormal eigenvectors of
$\mathbf{AA^{H}}$ and the diagonal matrix $\mathbf{\Lambda}$ has the singular
values of $\mathbf{A}$ on its main diagonal.
\begin{proposition}\label{pro:eigenv_compl}
    Suppose \emph{$\mathbf{x} \in \C^{n}$} is a complex-valued random vector with complementary covariance matrix \emph{$\mathbf{P}_{\mathbf{x}}\in {\C}^{n \times
    n}$}.
    Then, there exist a unitary matrix \emph{$\mathbf{Q}_{\mathbf{x}} \in {\C}^{n \times
    n}$} and a diagonal
    matrix \emph{$\mathbf{\Lambda}_{\mathbf{x}} \in {\R}^{n \times n}$} with non-negative entries,
    such that\emph{
    \begin{align*}
        \underline{\mathbf{P}}_{\mathbf{x}}=\overline{\mathbf{Q}}_{\mathbf{x}} \underline{\mathbf{\Lambda}}_{\mathbf{x}} {\overline{\mathbf{Q}}}_{\mathbf{x}}^{T}.
    \end{align*}}
    represents the eigenvalue decomposition of $\underline{\mathbf{P}}_{\mathbf{x}}$. The diagonal entries of \emph{$\mathbf{\Lambda}_{\mathbf{x}}$} are the
    singular values of \emph{$\hspace*{2mm}\mathbf{P}_{\mathbf{x}}$}. In particular,
    \emph{$\underline{\mathbf{\Lambda}}_{\mathbf{x}}=\mbox{diag}^{2n \times 2n}\left\{\mathbf{\Lambda}_{\mathbf{x}},-\mathbf{\Lambda}_{\mathbf{x}}
    \right\}$}.
\end{proposition}
\bproof Consider the Takagi factorization \eqref{takagi_equation} of the
symmetric $\mathbf{P}_{\mathbf{x}}$ and apply Lemma
\ref{lem:algebraic_props}, i.e.,
\begin{align*}
    \underline{\mathbf{P}}_{\mathbf{x}}=\underline{\mathbf{Q}_{\mathbf{x}} \left(\mathbf{\Lambda}_{\mathbf{x}} \mathbf{Q}_{\mathbf{x}}^{T}\right)} \overset{\eqref{alg_a}}{=}
    \overline{\mathbf{Q}}_{\mathbf{x}} \underline{\mathbf{\Lambda}_{\mathbf{x}} \mathbf{Q}_{\mathbf{x}}^{T}} =
    \overline{\mathbf{Q}}_{\mathbf{x}} \underline{\mathbf{\Lambda}_{\mathbf{x}} \left(\mathbf{Q}_{\mathbf{x}}^{H}\right)^{*}} \overset{\eqref{alg_b}}{=}
    \overline{\mathbf{Q}}_{\mathbf{x}} \underline{\mathbf{\Lambda}}_{\mathbf{x}} \overline{\mathbf{Q}_{\mathbf{x}}^{H}} \overset{\eqref{alg_c}}{=}
    \overline{\mathbf{Q}}_{\mathbf{x}} \underline{\mathbf{\Lambda}}_{\mathbf{x}} \overline{\mathbf{Q}}_{\mathbf{x}}^{T},
 \end{align*}
 which represents the eigenvalue decomposition of $\underline{\mathbf{P}}_{\mathbf{x}}$, since $\overline{\mathbf{Q}}_{\mathbf{x}}$ is orthonormal according to \eqref{alg_d}.\eproof
An interesting question that is directly related to complex-valued random
vectors is the characterization of the set of complementary covariance
matrices. For covariance matrices such a characterization is well known,
i.e., a matrix is a \emph{valid} covariance matrix, which means that there
exists a random vector with this covariance matrix, if and only if it is
Hermitian and non-negative definite. In order to obtain an analogous result
for complementary covariance matrices, we introduce the following
notion.\footnote{Cf. also the relation to the \emph{Karhunen-Lo\`{e}ve transform}
\cite{Karhu47,loeve48}, also known as \emph{Hotelling transform}
\cite{hotelling33}, and the \emph{Mahalanobis transform}, e.g.,
\cite{Eldar_IT03} and references therein.}
\begin{definition}\label{def:gen_chol}
    A matrix \emph{$\mathbf{B} \in {\C}^{n \times n}$} is said to be {\em generalized Cholesky factor} of
    a positive definite Hermitian matrix \emph{$\mathbf{A} \in {\C}^{n \times
    n}$}, if it satisfies \emph{$\mathbf{A}=\mathbf{B} \mathbf{B}^{H}$}.
\end{definition}
Since $\det\mathbf{A}=\left|\det\mathbf{B} \right|^{2}$, a generalized
Cholesky factor is always a non-singular matrix. Note that the conventional
Cholesky decomposition (cf. \cite{golub96}), $\mathbf{A=L L^{H}}$, where
$\mathbf{L}$ is lower-triangular, yields a generalized Cholesky factor
$\mathbf{L}$. But there are also other ways of constructing a generalized
Cholesky factor. Let $\mathbf{A=}$ $\mathbf{U D U^{H}}$ be the eigenvalue
decomposition of $\mathbf{A}$. For any matrix $\mathbf{T}$, which satisfies
$\mathbf{D=T T^{H}}$, $\mathbf{B=U T}$ is a generalized Cholesky factor.
Hence, a generalized Cholesky factor is not uniquely defined. However, we
have the following characterization.
\begin{proposition}
    Suppose \emph{$\mathbf{B}$} is a generalized Cholesky factor of
    \emph{$\mathbf{A}$}. Then, for any unitary matrix \emph{$\mathbf{U}$},
    \emph{$\mathbf{C=BU}$} is also a generalized Cholesky factor.
    Conversely, if \emph{$\mathbf{B}$} and \emph{$\mathbf{C}$} are generalized
    Cholesky factors, then, there exists a unitary matrix \emph{$\mathbf{U}$},
    such that \emph{$\mathbf{C=BU}$}.
\end{proposition}
\bproof For non-singular $\mathbf{B}$ and $\mathbf{C}$ we have
\begin{align*}
    \mathbf{B}\mathbf{B}^{H}=\mathbf{C}\mathbf{C}^{H}\quad\Leftrightarrow\quad\left(\mathbf{B}^{-1}\mathbf{C}\right)^{-1}=\left(\mathbf{B}^{-1}\mathbf{C}\right)^{H},
\end{align*}
which implies both statements.\eproof The next theorem presents the promised
criterion for a matrix to be a  complementary covariance matrix. More
precisely, it is a criterion in terms of both covariance matrix and
complementary covariance matrix. We will call $\{\mathbf{C},\mathbf{P}\}$ a
\emph{valid pair of covariance matrix and complementary covariance matrix},
if there exists a complex-valued random vector with covariance matrix
$\mathbf{C}$ and complementary covariance matrix $\mathbf{P}$.
\begin{theorem}\label{pseud_criterion}
     Suppose \emph{$\mathbf{C} \in {\C}^{ n \times n}$} is non-singular and \emph{$\mathbf{P} \in {\C}^{ n \times n}$}. Then, \emph{$\{\mathbf{C},\mathbf{P}\}$} is a
     valid pair of covariance matrix and complementary covariance matrix if
     and only if \emph{$\mathbf{C}$} is Hermitian and non-negative definite, \emph{$\mathbf{P}$} is symmetric,
     and the singular values of \emph{$\mathbf{B}^{-1}\mathbf{P} \mathbf{B}^{-T}$} are smaller or equal to
     \emph{$1$}, where \emph{$\mathbf{B}$} denotes an arbitrary generalized Cholesky factor of \emph{$\mathbf{C}$}.
\end{theorem}
\bproof The requirements that $\mathbf{C}$ is Hermitian and non-negative
definite as well as that $\mathbf{P}$ is symmetric are obvious. Furthermore,
observe that the singular values of $\mathbf{B}^{-1}\mathbf{P}
\mathbf{B}^{-T}$ do not depend on the choice of the generalized Cholesky
factor $\mathbf{B}$.

Suppose we are given a complex-valued random vector $\mathbf{x} \in \C^{n}$
with covariance matrix $\mathbf{C}$ and complementary covariance matrix
$\mathbf{P}$. Consider the random vector
$\mathbf{y}\triangleq\mathbf{B}^{-1}\mathbf{x}$. Clearly,
$\mathbf{C}_{\mathbf{y}}=\mathbf{I}_{n}$ and
$\mathbf{P}_{\mathbf{y}}=\mathbf{B}^{-1}\mathbf{P} \mathbf{B}^{-T}$. From
\eqref{expr_cov_compl}, i.e., $\mathbf{C}_{\mathbf{y}^{\text{(r)}}} =
\frac{1}{2} \left(\mathbf{I}_{2n} + \underline{\mathbf{B}^{-1}\mathbf{P}
\mathbf{B}^{-T}}\right)$, and the fact that
$\mathbf{C}_{\mathbf{y}^{\text{(r)}}}$ is non-negative definite, we conclude
with Proposition \ref{pro:eigenv_compl}, that the singular values of
$\mathbf{B}^{-1}\mathbf{P} \mathbf{B}^{-T}$ are smaller or equal to $1$.

Conversely, consider a complex-valued random vector $\mathbf{y}$, e.g., a
Gaussian distributed one, defined by the covariance matrix of its real
representation as $\mathbf{C}_{\mathbf{y}^{\text{(r)}}} \triangleq
\frac{1}{2} \left(\mathbf{I}_{2n} + \underline{\mathbf{B}^{-1}\mathbf{P}
\mathbf{B}^{-T}}\right)$. According to Proposition \ref{pro:eigenv_compl},
such a random vector exists, since $\mathbf{C}_{\mathbf{y}^{\text{(r)}}}$ is
Hermitian and non-negative definite provided that the singular values of
$\mathbf{B}^{-1}\mathbf{P} \mathbf{B}^{-T}$ are smaller or equal to $1$. It
has covariance matrix $\mathbf{C}_{\mathbf{y}}=\mathbf{I}_{n}$ and
complementary covariance matrix
$\mathbf{P}_{\mathbf{y}}=\mathbf{B}^{-1}\mathbf{P} \mathbf{B}^{-T}$, cf.
\eqref{expr_cov_compl}. Then, the random vector $\mathbf{x}\triangleq
\mathbf{B} \mathbf{y}$ has covariance matrix
$\mathbf{C}_{\mathbf{y}}=\mathbf{C}$ and complementary covariance matrix
$\mathbf{P}_{\mathbf{y}}=\mathbf{P}$.\eproof {\bf Remarks.} Apparently, the
importance of the singular values of $\mathbf{B}^{-1}\mathbf{P}
\mathbf{B}^{-T}$ in the context of complex-valued random vectors was first
observed in \cite{Tauboeck02} (for the above criterion and a generalized
maximum entropy theorem) and independently in \cite{Schreier02ICASSP}, where
they were introduced as \emph{canonical coordinates}
\cite{hotelling33,hotelling36,scharf_sp98,scharf_sp00,schreier_spl06} between
a complex-valued random vector and its complex conjugate. Note that in
\cite{Schreier02ICASSP}, the matrix $\mathbf{B}^{-1}\mathbf{P}
\mathbf{B}^{-T}$ is called \emph{coherence matrix} between a random vector
and its complex conjugate. Interestingly, the approach of
\cite{Schreier02ICASSP} differs from the approach taken here (and taken in
\cite{Tauboeck02}) in that \cite{Schreier02ICASSP} employs a complex-valued
\emph{augmented algebra} to study second-order properties of complex-valued
random vectors, whereas \eqref{matr_real_repr} introduces a real-valued
representation into real and imaginary parts. Later, in \cite{Koivunen_IT06},
the singular values were also termed \emph{circularity coefficients} and the
whole set of singular values was referred to as \emph{circularity spectrum}.
We also note that the condition of Theorem \ref{pseud_criterion} on the
singular values, can be equivalently expressed in terms of the Euclidean
operator norm $\| \cdot \|_{2}$ as $\left\|\mathbf{B}^{-1}\mathbf{P}
\mathbf{B}^{-T}\right\|_{2} \leq 1$.

\section{Circular Analog of a Complex-Valued Random Vector}\label{sec:analogs}
In this section we consider the following problem: suppose we are given a
complex-valued random vector, which is non-circular. Can we find a random
vector, which is as ``similar'' as possible to the original random vector but
circular instead? Obviously, this depends on what is meant by ``similar'' and
is, therefore, mainly a matter of definition. However, if we can show useful
properties and/or theorems with this circularized random vector, its
introduction is reasonable. Our approach for associating a circular random
vector to a (possibly) non-circular one is motivated by the well-known method
used for stationarizing a cyclostationary random process \cite{papoulis91}.
\begin{definition}\label{def:a1}
    Suppose \emph{$\mathbf{x} \in \C^{n}$} is a complex-valued random vector.
    Then, the random vector \emph{$\mathbf{x}_{\text{(a)}} \triangleq e^{\jmath 2
    \pi \psi} \mathbf{x}$}, where \emph{$\psi \in [0,1[$} is a uniformly distributed random variable
    independent of
    \emph{$\mathbf{x}$}, is said to be \emph{circular analog}
    of \emph{$\mathbf{x}$}.
\end{definition}
In the following, we will show that the circular analog is indeed a circular
random vector. The next lemma expresses the distribution of
$\mathbf{x}_{\text{(a)}}$ in terms of the distribution of $\mathbf{x}$ (for
both polar and sheared-polar representations).
\begin{proposition}\label{pro:a1_main}
    Suppose \emph{$\mathbf{x} \in \C^{n}$} is a complex-valued random vector.
    Then, the pdfs of the polar representations and sheared-polar representations of \emph{$\mathbf{x}$} and
    its circular analog \emph{$\mathbf{x}_{\text{(a)}}$} are related according to
    \emph{
\begin{align}\label{a1_polar}
    f_{\mathbf{x}^{\text{(p)}}_{\text{(a)}}}(r_1,\ldots,r_n,\phi_1,\ldots,\phi_n) &= \int_{0}^{1}f_{\mathbf{x}^{\text{(p)}}}\left(r_1,\ldots,r_n,[\phi_1-\varphi]_{[0,1[},\ldots,[\phi_n-\varphi]_{[0,1[}\right)d\varphi \quad \lambda_{2n}\text{-a.e.},\\\label{a1_sheared}
    f_{\mathbf{x}^{\text{(s)}}_{\text{(a)}}}(r_1,\ldots,r_n,\phi_1,\ldots,\phi_n) &= \int_{0}^{1} f_{\mathbf{x}^{\text{(s)}}}\left(r_1,\ldots,r_n,\phi_1, \phi_2,\ldots,\phi_n\right)d\phi_{n}\quad \lambda_{2n}\text{-a.e.},
\end{align}}
respectively.
\end{proposition}
\bproof For \eqref{a1_polar}, consider the joint pdf of
$\mathbf{x}^{\text{(p)}}_{\text{(a)}}$ and $\psi$, i.e.,
$f_{\mathbf{x}^{\text{(p)}}_{\text{(a)}};\psi}(r_1,\ldots,r_n,\phi_1,\ldots,\phi_n,\varphi)
=f_{\mathbf{x}^{\text{(p)}}_{\text{(a)}}|\psi}(r_1,\ldots$
$\ldots,r_n,\phi_1,\ldots,\phi_n|\varphi)\,f_{\psi}(\varphi)=f_{\mathbf{x}^{\text{(p)}}}\left(r_1,\ldots,r_n,[\phi_1-\varphi]_{[0,1[},\ldots,[\phi_n-\varphi]_{[0,1[}\right)$
and marginalize with respect to $\varphi$. \eqref{a1_sheared} follows from
\eqref{a1_polar} using Lemma \ref{lem:sheared_distr} and identity
\eqref{mod_identity}.\eproof Observe that
$f_{\mathbf{x}^{\text{(s)}}_{\text{(a)}}}$ does not depend on
$\phi_{n}$\hspace*{1mm}$\lambda_{2n}\text{-a.e.}$, so that Corollary
\ref{cor:circ_sheared} implies circularity of $\mathbf{x}_{\text{(a)}}$.

\subsection{Divergence Characterization}
Here, we present a characterization of the circular analog of a
complex-valued random vector that further supports the chosen definition. It
is based on the \emph{Kullback-Leibler divergence} (or \emph{relative
entropy}) \cite{Cover91,Dembo10}, which can be regarded as a distance measure
between two probability measures. For complex-valued random vectors, whose
real representations are distributed according to multivariate pdfs, the
Kullback-Leibler divergence $D(\mathbf{x}\|\mathbf{y})$ between $\mathbf{x}
\in \C^{n}$ and $\mathbf{y} \in \C^{n}$ is defined as
\begin{align*}
    D\big(\mathbf{x}\|\mathbf{y}\big) \triangleq D\big(\mathbf{x}^{\text{(r)}}\|\mathbf{y}^{\text{(r)}}\big) =
    \int\limits_{\R^{2n}} f_{\mathbf{x}^{\text{(r)}}}(\xxi) \log \frac{f_{\mathbf{x}^{\text{(r)}}}(\xxi)}{f_{\mathbf{y}^{\text{(r)}}}(\xxi)}d\xxi \quad \in \R_{0}^{+} \cup \{\infty \},
\end{align*}
where we set $0 \log 0 \triangleq 0$ and $0 \log \frac{0}{0} \triangleq 0$
(motivated by continuity). Here, $D(\mathbf{x}\|\mathbf{y})$ is finite only
if the support set of $f_{\mathbf{x}^{\text{(r)}}}$ is contained in the
support set of
$f_{\mathbf{y}^{\text{(r)}}}$\hspace*{1mm}$\lambda_{2n}\text{-a.e.}$. Note
that $D(\mathbf{x}\|\mathbf{y})=0$ if and only if
$f_{\mathbf{x}^{\text{(r)}}} =
f_{\mathbf{y}^{\text{(r)}}}$\hspace*{1mm}$\lambda_{2n}\text{-a.e.}$
\cite{Cover91}. The next lemma shows that $D(\mathbf{x}\|\mathbf{y})$ can be
equivalently expressed in terms of polar and sheared-polar representations.
\begin{lemma}\label{KL_compute}
Suppose \emph{$\mathbf{x} \in \C^{n}$} and \emph{$\mathbf{y} \in \C^{n}$} are
complex-valued random vectors. Then, the Kullback-Leibler divergence
\emph{$D(\mathbf{x}\|\mathbf{y})$} can be computed from the respective polar
and sheared-polar representations of \emph{$\mathbf{x}$} and
\emph{$\mathbf{y}$} according to \emph{
\begin{align*}
    D\big(\mathbf{x}\|\mathbf{y}\big) &= D\big(\mathbf{x}^{\text{(p)}}\|\mathbf{y}^{\text{(p)}}\big) = \hspace*{-13mm}\int\limits_{\hspace*{10mm}(\R_{0}^{+})^{n}\times ([0,1[)^{n}} \hspace*{-14mm} f_{\mathbf{x}^{\text{(p)}}}(\xxi) \log
    \frac{f_{\mathbf{x}^{\text{(p)}}}(\xxi)}{f_{\mathbf{y}^{\text{(p)}}}(\xxi)}d\xxi\\[2mm]
    &= D\big(\mathbf{x}^{\text{(s)}}\|\mathbf{y}^{\text{(s)}}\big) = \hspace*{-13mm}\int\limits_{\hspace*{10mm}(\R_{0}^{+})^{n}\times ([0,1[)^{n}} \hspace*{-14mm} f_{\mathbf{x}^{\text{(s)}}}(\xxi) \log
    \frac{f_{\mathbf{x}^{\text{(s)}}}(\xxi)}{f_{\mathbf{y}^{\text{(s)}}}(\xxi)}d\xxi.
\end{align*}}
\end{lemma}
\bproof With $\mathcal{A}\triangleq (\R_{0}^{+})^{n}\times ([0,1[)^{n}$,
\begin{align*}
    D\big(\mathbf{x}^{\text{(r)}}\|\mathbf{y}^{\text{(r)}}\big) &= \hspace*{-8mm}
    \int\limits_{\hspace*{6mm} \mathbb{T}^{\text{(p$\rightarrow$\,r)}}(\mathcal{A})} \hspace*{-9mm} f_{\mathbf{x}^{\text{(r)}}}(\xxi) \log
    \frac{f_{\mathbf{x}^{\text{(r)}}}(\xxi)}{f_{\mathbf{y}^{\text{(r)}}}(\xxi)}d\xxi
    = \hspace*{0mm}
    \int\limits_{\hspace*{0mm} \mathcal{A}} \hspace*{0mm} f_{\mathbf{x}^{\text{(r)}}}\left(\mathbb{T}^{\text{(p$\rightarrow$\,r)}}(\xxi)\right) \log
    \frac{f_{\mathbf{x}^{\text{(r)}}}\left(\mathbb{T}^{\text{(p$\rightarrow$\,r)}}(\xxi)\right)}{f_{\mathbf{y}^{\text{(r)}}}\left(\mathbb{T}^{\text{(p$\rightarrow$\,r)}}(\xxi)\right)} \left|J_{\mathbb{T}^{\text{(p$\rightarrow$\,r)}}}(\xxi) \right|d\xxi\\
    &= \hspace*{0mm}
    \int\limits_{\hspace*{0mm} \mathcal{A}} \hspace*{0mm} f_{\mathbf{x}^{\text{(p)}}}(\xxi) \log
    \frac{f_{\mathbf{x}^{\text{(p)}}}(\xxi)}{f_{\mathbf{y}^{\text{(p)}}}(\xxi)}d\xxi = D\big(\mathbf{x}^{\text{(p)}}\|\mathbf{y}^{\text{(p)}}\big),
\end{align*}
where Lemma \ref{lem:polar_distr} has been used. Furthermore, using the
mapping $\widetilde{\mathbb{T}}^{\text{(s$\rightarrow$\,p)}}$ and the
appropriate partition $\{\mathcal{A}_1,\ldots,\mathcal{A}_N\}$ of
$\mathcal{A}$, cf. the proof of Lemma \ref{lem:sheared_distr},
\begin{align*}
    D\big(\mathbf{x}^{\text{(p)}}\|\mathbf{y}^{\text{(p)}}\big) &= \hspace*{-8mm}
    \int\limits_{\hspace*{6mm} \mathbb{T}^{\text{(s$\rightarrow$\,p)}}(\mathcal{A})} \hspace*{-9mm} f_{\mathbf{x}^{\text{(p)}}}(\xxi) \log
    \frac{f_{\mathbf{x}^{\text{(p)}}}(\xxi)}{f_{\mathbf{y}^{\text{(p)}}}(\xxi)}d\xxi
    = \hspace*{-19mm} \int\limits_{\hspace*{17mm} \widetilde{\mathbb{T}}^{\text{(p$\rightarrow$\,s})}\left(\mathbb{T}^{\text{(s$\rightarrow$\,p)}}(\mathcal{A})\right)} \hspace*{-21mm}
    f_{\mathbf{x}^{\text{(p)}}}\left(\widetilde{\mathbb{T}}^{\text{(s$\rightarrow$\,p)}}(\xxi)\right) \log
    \frac{f_{\mathbf{x}^{\text{(p)}}}\left(\widetilde{\mathbb{T}}^{\text{(s$\rightarrow$\,p)}}(\xxi)\right)}{f_{\mathbf{y}^{\text{(p)}}}\left(\widetilde{\mathbb{T}}^{\text{(s$\rightarrow$\,p)}}(\xxi)\right)}d\xxi\\
    &= \bigcup\limits_{i=1}^{N}\hspace*{0mm}\int\limits_{\mathcal{A}_i} \hspace*{0mm}
    f_{\mathbf{x}^{\text{(p)}}}\left(\widetilde{\mathbb{T}}^{\text{(s$\rightarrow$\,p)}}(\xxi+\mathbf{k}_i)\right) \log
    \frac{f_{\mathbf{x}^{\text{(p)}}}\left(\widetilde{\mathbb{T}}^{\text{(s$\rightarrow$\,p)}}(\xxi+\mathbf{k}_i)\right)}{f_{\mathbf{y}^{\text{(p)}}}\left(\widetilde{\mathbb{T}}^{\text{(s$\rightarrow$\,p)}}(\xxi+\mathbf{k}_i)\right)}d\xxi\\
    &= \bigcup\limits_{i=1}^{N}\hspace*{0mm}\int\limits_{\mathcal{A}_i} \hspace*{0mm}
    f_{\mathbf{x}^{\text{(p)}}}\left(\mathbb{T}^{\text{(s$\rightarrow$\,p)}}(\xxi)\right) \log
    \frac{f_{\mathbf{x}^{\text{(p)}}}\left(\mathbb{T}^{\text{(s$\rightarrow$\,p)}}(\xxi)\right)}{f_{\mathbf{y}^{\text{(p)}}}\left(\mathbb{T}^{\text{(s$\rightarrow$\,p)}}(\xxi)\right)}d\xxi = D\big(\mathbf{x}^{\text{(s)}}\|\mathbf{y}^{\text{(s)}}\big),
\end{align*}
where Lemma \ref{lem:sheared_distr} has been used.\eproof We intend to prove
a theorem, which states that the circular analog has a smaller ``distance''
from the given complex-valued random vector than any other circular random
vector. To that end, consider the sheared-polar representation of
$\mathbf{x}$, i.e., $\mathbf{x}^{\text{(s)}} \in \R^{2n}$, and form the
``reduced'' vector $\tilde{\mathbf{x}}^{\text{(s)}} \in \R^{2n-1}$ by only
taking the first $2n-1$ elements of $\mathbf{x}^{\text{(s)}}$. Clearly, its
pdf is given by marginalization, i.e.,
\begin{align*}
    f_{\tilde{\mathbf{x}}^{\text{(s)}}}(\tilde{\xxi}) = \int_{0}^{1} f_{\mathbf{x}^{\text{(s)}}}(r_1,\ldots,r_n,\phi_1,\ldots,\phi_n) d \phi_n, \quad \text{where} \quad\tilde{\xxi}\triangleq(r_1,\ldots,r_n,\phi_1,\ldots,\phi_{n-1}).
\end{align*}
Furthermore, let $\widetilde{\mathcal{S}}_{\mathbf{x}} \subset
(\R_{0}^{+})^{n}\times ([0,1[)^{n-1}$ denote the support set of
$f_{\tilde{\mathbf{x}}^{\text{(s)}}}$. Note that
$f_{\tilde{\mathbf{x}}^{\text{(s)}}}(\tilde{\xxi})=0$ is equivalent to
$f_{\mathbf{x}^{\text{(s)}}}(\tilde{\xxi},\phi_n) =
0$\hspace*{1mm}$\lambda_{1}\text{-a.e.}$ (for fixed $\tilde{\xxi}$). We have,
\begin{align}\label{cond_dens}
    f_{\mathbf{x}^{\text{(s)}}}(\tilde{\xxi},\phi_n)=f_{\tilde{\mathbf{x}}^{\text{(s)}}}(\tilde{\xxi})\, f_{\vartheta|\tilde{\mathbf{x}}^{\text{(s)}}}(\phi_n|\tilde{\xxi}),  \quad \tilde{\xxi} \in \widetilde{\mathcal{S}}_{\mathbf{x}},\,\, \phi_n \in [0,1[,
\end{align}
where $\vartheta \triangleq \left(\mathbf{x}^{\text{(s)}}\right)_{2n}$ is the
last element of $\mathbf{x}^{\text{(s)}}$.
\begin{theorem}\label{a1_charac_div}
    Suppose \emph{$\mathbf{x} \in \C^{n}$} is a complex-valued random vector.
    Then, a circular random vector \emph{$\mathbf{y} \in \C^{n}$}  is the circular analog
    of \emph{$\mathbf{x}$}, i.e., \emph{$\mathbf{y}=\mathbf{x}_{\text{(a)}} $}, if and only if  
    it minimizes the
    Kullback-Leibler divergence to \emph{$\mathbf{x} \in \C^{n}$} within the whole set of circular random vectors, i.e., if and only if \emph{
    \begin{align*}
        D\big(\mathbf{x}\|\mathbf{y}\big) = \inf_{\mathbf{c} \in \mathcal{C}_{n}} D\big(\mathbf{x}\|\mathbf{c}\big).
    \end{align*}}
    Furthermore, \emph{
    \begin{align*}
        D\big(\mathbf{x}\|\mathbf{x}_{\text{(a)}}\big) = \inf_{\mathbf{c} \in \mathcal{C}_{n}} D\big(\mathbf{x}\|\mathbf{c}\big)= \int\limits_{\hspace*{0mm}\widetilde{\mathcal{S}}_{\mathbf{x}}} \hspace*{0mm} f_{\tilde{\mathbf{x}}^{\text{(s)}}}(\tilde{\xxi})\left(\int_{0}^{1}f_{\vartheta|\tilde{\mathbf{x}}^{\text{(s)}}}(\phi_n|\tilde{\xxi}) \log
    f_{\vartheta|\tilde{\mathbf{x}}^{\text{(s)}}}(\phi_n|\tilde{\xxi})d\phi_n\right) d\tilde{\xxi} \triangleq h(\vartheta|\tilde{\mathbf{x}}^{\text{(s)}}),
    \end{align*}}
    where \emph{$h(\vartheta|\tilde{\mathbf{x}}^{\text{(s)}})$} denotes the \emph{conditional differential
    entropy} of \emph{$\vartheta$}
    given
    \emph{$\tilde{\mathbf{x}}^{\text{(s)}}$}, cf. \emph{\cite{Cover91}} and Definition \ref{def:cond_entr}, with \emph{$\vartheta$}
    and
    \emph{$\tilde{\mathbf{x}}^{\text{(s)}}$} according to
    \eqref{cond_dens}.
\end{theorem}
\bproof Suppose $\mathbf{c}\in\mathcal{C}_{n}$ and consider its sheared-polar
representation $\mathbf{c}^{\text{(s)}} \in \R^{2n}$. Due to the circularity
of $\mathbf{c}$, $f_{\mathbf{c}^{\text{(s)}}}(\xxi) =
f_{\mathbf{c}^{\text{(s)}}}(\tilde{\xxi})$\hspace*{1mm}$\lambda_{2n}\text{-a.e.}$,
and, according to Lemma \ref{KL_compute},
\begin{align*}
    D\big(\mathbf{x}\|\mathbf{c}\big) &= \hspace*{-5mm}\int\limits_{\hspace*{3mm}\widetilde{\mathcal{S}}_{\mathbf{x}}\times [0,1[} \hspace*{-6mm} {f_{\tilde{\mathbf{x}}^{\text{(s)}}}(\tilde{\xxi})\, f_{\vartheta|\tilde{\mathbf{x}}^{\text{(s)}}}(\phi_n|\tilde{\xxi})} \log
    \frac{f_{\tilde{\mathbf{x}}^{\text{(s)}}}(\tilde{\xxi})\, f_{\vartheta|\tilde{\mathbf{x}}^{\text{(s)}}}(\phi_n|\tilde{\xxi})}{f_{\mathbf{c}^{\text{(s)}}}(\tilde{\xxi})}d\tilde{\xxi} d\phi_n\\
    &\overset{(*)}{=}\hspace*{0mm}\int\limits_{\hspace*{0mm}\widetilde{\mathcal{S}}_{\mathbf{x}}} \hspace*{0mm} f_{\tilde{\mathbf{x}}^{\text{(s)}}}(\tilde{\xxi}) \log
    \frac{f_{\tilde{\mathbf{x}}^{\text{(s)}}}(\tilde{\xxi})}{f_{\mathbf{c}^{\text{(s)}}}(\tilde{\xxi})}d\tilde{\xxi}
    +\hspace*{0mm}\int\limits_{\hspace*{0mm}\widetilde{\mathcal{S}}_{\mathbf{x}}} \hspace*{0mm} f_{\tilde{\mathbf{x}}^{\text{(s)}}}(\tilde{\xxi})\left(\int_{0}^{1}f_{\vartheta|\tilde{\mathbf{x}}^{\text{(s)}}}(\phi_n|\tilde{\xxi}) \log
    f_{\vartheta|\tilde{\mathbf{x}}^{\text{(s)}}}(\phi_n|\tilde{\xxi})d\phi_n\right) d\tilde{\xxi}\\
    &= D\big(\tilde{\mathbf{x}}^{\text{(s)}}\|\tilde{\mathbf{c}}^{\text{(s)}}\big)
    +\hspace*{0mm}\int\limits_{\hspace*{0mm}\widetilde{\mathcal{S}}_{\mathbf{x}}} \hspace*{0mm} f_{\tilde{\mathbf{x}}^{\text{(s)}}}(\tilde{\xxi})\left(\int_{0}^{1}f_{\vartheta|\tilde{\mathbf{x}}^{\text{(s)}}}(\phi_n|\tilde{\xxi}) \log
    f_{\vartheta|\tilde{\mathbf{x}}^{\text{(s)}}}(\phi_n|\tilde{\xxi})d\phi_n\right) d\tilde{\xxi},
\end{align*}
where $\tilde{\mathbf{c}}^{\text{(s)}} \in \R^{2n-1}$ is the corresponding
``reduced'' vector of $\mathbf{c}^{\text{(s)}}$. For the validity of $(*)$,
we also refer to \cite[Theorem D.13]{Dembo10}. It follows that
\begin{align}\label{div_infimum}
  \inf_{\mathbf{c} \in \mathcal{C}_{n}} D\big(\mathbf{x}\|\mathbf{c}\big) = \int\limits_{\hspace*{0mm}\widetilde{\mathcal{S}}_{\mathbf{x}}} \hspace*{0mm} f_{\tilde{\mathbf{x}}^{\text{(s)}}}(\tilde{\xxi})\left(\int_{0}^{1}f_{\vartheta|\tilde{\mathbf{x}}^{\text{(s)}}}(\phi_n|\tilde{\xxi}) \log
    f_{\vartheta|\tilde{\mathbf{x}}^{\text{(s)}}}(\phi_n|\tilde{\xxi})d\phi_n\right) d\tilde{\xxi},
\end{align}
and the infimum is achieved for
$f_{\tilde{\mathbf{c}}^{\text{(s)}}}=f_{\tilde{\mathbf{x}}^{\text{(s)}}}$\hspace*{1mm}$\lambda_{2n-1}\text{-a.e.}$.
Since for the circular analog $\mathbf{x}_{\text{(a)}}$ of $\mathbf{x}$,
$f_{\mathbf{x}^{\text{(s)}}_{\text{(a)}}}(\xxi)=f_{\tilde{\mathbf{x}}^{\text{(s)}}}(\tilde{\xxi})$\hspace*{1mm}$\lambda_{2n}\text{-a.e.}$,
and since
$f_{\mathbf{c}^{\text{(s)}}}(\xxi)=f_{\tilde{\mathbf{c}}^{\text{(s)}}}(\tilde{\xxi})$\hspace*{1mm}$\lambda_{2n}\text{-a.e.}$,
the infimum is achieved if and only if
$f_{\mathbf{c}^{\text{(s)}}}=f_{\mathbf{x}^{\text{(s)}}_{\text{(a)}}}$\hspace*{1mm}$\lambda_{2n}\text{-a.e.}$,
i.e., $\mathbf{c}=\mathbf{x}_{\text{(a)}}$.\eproof

\subsection{Complex-Valued Random Vectors with Finite Second-Order Moments}
In this section, we establish important properties of the circular analog
$\mathbf{x}_{\text{(a)}}$ of a complex-valued random vector $\mathbf{x}$,
whose second-order moments exist. Clearly, both mean vector and complementary
covariance matrix of $\mathbf{x}_{\text{(a)}}$ are vanishing. For the
covariance matrix, we have the following result.
\begin{theorem}\label{a1_covmat} Suppose \emph{$\mathbf{x} \in \C^{n}$} is a zero-mean complex-valued random vector with finite second-order
moments. Then, the covariance matrix of the circular analog
\emph{$\mathbf{x}_{\text{(a)}}$} equals the covariance matrix of
\emph{$\mathbf{x}$}, i.e.,
\emph{$\mathbf{C}_{\mathbf{x}_{\text{(a)}}}=\mathbf{C}_{\mathbf{x}}$}.
\end{theorem}
\bproof For the correlation between the $k$th and $l$th entry of
$\mathbf{x}_{\text{(a)}}$,
\begin{align*}
 \E{\left(\mathbf{x}_{\text{(a)}}\right)_{k} \left(\mathbf{x}_{\text{(a)}}\right)_{l}^{*}}&= \int_{\R^{2n}} (\xi_k+\jmath\xi_{k+n}) (\xi_l-\jmath\xi_{l+n}) f_{\mathbf{x}_{\text{(a)}}^{\text{(r)}}}(\xxi)d\xxi\\
 &= \int_{\R^{2n}} \int_{0}^{1}(\xi_k+\jmath\xi_{k+n}) (\xi_l-\jmath\xi_{l+n}) f_{\mathbf{x}^{\text{(r)}}|\psi}(\xxi|\varphi) d\varphi d\xxi\\
 &\overset{(*)}{=} \int_{0}^{1} \int_{\R^{2n}} (\xi_k+\jmath\xi_{k+n}) (\xi_l-\jmath\xi_{l+n}) f_{\mathbf{x}^{\text{(r)}}|\psi}(\xxi|\varphi) d\xxi d\varphi\\
 &= \int_{0}^{1} \E{\left(e^{\jmath 2 \pi \varphi}\mathbf{x}\right)_{k} \left(e^{\jmath 2 \pi \varphi}\mathbf{x}\right)_{l}^{*}}d\varphi = \E{\left(\mathbf{x}\right)_{k} \left(\mathbf{x}\right)_{l}^{*}},
\end{align*}
where $\psi$ denotes the uniformly distributed random variable used for
defining ${\mathbf{x}}_{\text{(a)}}$ (see Definition \ref{def:a1}) and $(*)$
follows from Fubini's Theorem \cite{Rud87}. \eproof The following theorem
states that the circular analog of an improper Gaussian distributed random
vector is non-Gaussian.
\begin{theorem}\label{a1_distr_gauss}
    Suppose \emph{$\mathbf{x} \in \C^{n}$} is a zero-mean, complex-valued, and  Gaussian distributed random
    vector with \emph{$\|\mathbf{B}_{\mathbf{x}}^{-1}\mathbf{P}_{\mathbf{x}}$ $\mathbf{B}_{\mathbf{x}}^{-T}\|_{2}<1$}
    such that its circular analog
    \emph{$\mathbf{x}_{\text{(a)}}$} is Gaussian distributed. Here, $\emph{$\mathbf{B}_{\mathbf{x}}$}$
    denotes a generalized Cholesky factor of the covariance matrix \emph{$\mathbf{C}_{\mathbf{x}}$} of \emph{$\mathbf{x}$} and \emph{$\mathbf{P}_{\mathbf{x}}$} denotes the
    complementary covariance matrix of \emph{$\mathbf{x}$}.  Then,
    \emph{$\mathbf{x}$} is proper.
\end{theorem}
\bproof We first prove the theorem for the special case
$\mathbf{C}_{\mathbf{x}}=\mathbf{I}_{n}$ and
$\mathbf{P}_{\mathbf{x}}=\mathbf{\Lambda}_{\mathbf{x}}$, where
$\mathbf{\Lambda}_{\mathbf{x}} \in {\R}^{n \times n}$ denotes a diagonal
matrix with non-negative diagonal entries $\lambda_i < 1$. For fixed
(deterministic) $\theta$, consider the random vector $\mathbf{y}_{(\theta)}
\triangleq e^{\jmath 2 \pi \theta}\mathbf{x}$, which has covariance matrix
$\mathbf{C}_{\mathbf{y}_{(\theta)}}=\mathbf{I}_{n}$ and complementary
covariance matrix $\mathbf{P}_{\mathbf{y}_{(\theta)}}=e^{\jmath 4 \pi \theta}
\mathbf{\Lambda}_{\mathbf{x}}$. According to \eqref{expr_cov_compl}, the
covariance matrix of its real representation
$\mathbf{y}_{(\theta)}^{\text{(r)}}$  is given by
\begin{align*}
  \mathbf{C}_{\mathbf{y}_{(\theta)}^{\text{(r)}}} = \frac{1}{2}\left[
                                                                \begin{array}{cc}
                                                                  \mathbf{I}_{n} & \mathbf{0} \\
                                                                  \mathbf{0} & \mathbf{I}_{n} \\
                                                                \end{array}
                                                              \right]+
                                                              \frac{1}{2}\left[
                                                                \begin{array}{lr}
                                                                  \cos (4\pi \theta) \,  \mathbf{\Lambda}_{\mathbf{x}} & \sin (4\pi \theta) \, \mathbf{\Lambda}_{\mathbf{x}} \\
                                                                  \sin (4\pi \theta) \, \mathbf{\Lambda}_{\mathbf{x}} & -\cos (4\pi \theta) \, \mathbf{\Lambda}_{\mathbf{x}} \\
                                                                \end{array}
                                                              \right],
\end{align*}
whose determinant is easily computed as
\begin{align*}
    \det\mathbf{C}_{\mathbf{y}_{(\theta)}^{\text{(r)}}}= 2^{-2n}
    \prod\limits_{i=1}^{n} (1-\lambda_{i}^{2}).
\end{align*}
Furthermore, its inverse is calculated as
\begin{align*}
  \mathbf{C}_{\mathbf{y}_{(\theta)}^{\text{(r)}}} = 2\left[
                                                                \begin{array}{cc}
                                                                  \left(\mathbf{I}_{n}- \mathbf{\Lambda}_{\mathbf{x}}^{2}\right)^{-1}& \mathbf{0} \\
                                                                  \mathbf{0} & \left(\mathbf{I}_{n}- \mathbf{\Lambda}_{\mathbf{x}}^{2}\right)^{-1} \\
                                                                \end{array}
                                                              \right]-
                                                              2\left[
                                                                \begin{array}{lr}
                                                                  \cos (4\pi \theta) \,  \mathbf{D}_{\mathbf{x}} & \sin (4\pi \theta) \, \mathbf{D}_{\mathbf{x}} \\
                                                                  \sin (4\pi \theta) \, \mathbf{D}_{\mathbf{x}} & -\cos (4\pi \theta) \, \mathbf{D}_{\mathbf{x}}\\
                                                                \end{array}
                                                              \right],
\end{align*}
where $\mathbf{D}_{\mathbf{x}} \triangleq  \mathbf{\Lambda}_{\mathbf{x}}
\left(\mathbf{I}_{n}- \mathbf{\Lambda}_{\mathbf{x}}^{2}\right)^{-1}$.
Therefore, the pdf of $\mathbf{y}_{(\theta)}^{\text{(r)}}$ is given by
\begin{align*}
    f_{\mathbf{y}^{\text{(r)}}_{(\theta)}}(\xxi) & = \frac{1}{\pi^{n} \prod\limits_{i=1}^{n} \sqrt{1-\lambda_{i}^{2}}} \exp\left(-\xxi^{T} \left[
                                                                \begin{array}{cc}
                                                                  \left(\mathbf{I}_{n}- \mathbf{\Lambda}_{\mathbf{x}}^{2}\right)^{-1}& \mathbf{0} \\
                                                                  \mathbf{0} & \left(\mathbf{I}_{n}- \mathbf{\Lambda}_{\mathbf{x}}^{2}\right)^{-1} \\
                                                                \end{array}
                                                              \right] \xxi  \right) \times \\
    & \hspace*{40mm} \times \exp\left(\xxi^{T} \left[
                                                                \begin{array}{lr}
                                                                  \cos (4\pi \theta) \,  \mathbf{D}_{\mathbf{x}} & \sin (4\pi \theta) \, \mathbf{D}_{\mathbf{x}} \\
                                                                  \sin (4\pi \theta) \, \mathbf{D}_{\mathbf{x}} & -\cos (4\pi \theta) \, \mathbf{D}_{\mathbf{x}}\\
                                                                \end{array}
                                                              \right]     \xxi  \right)  \quad \lambda_{2n}\text{-a.e.}.
\end{align*}
Since $f_{\mathbf{x}_{\text{(a)}}^{\text{(r)}}}(\xxi) =
\int_{0}^{1}f_{\mathbf{y}^{\text{(r)}}_{(\theta)}}(\xxi)
d\theta$\hspace*{2mm}$\lambda_{2n}\text{-a.e.}$, we obtain
\begin{align}\label{a1_pdf_gauss}
    f_{\mathbf{x}_{\text{(a)}}^{\text{(r)}}}(\xxi) & = \frac{1}{\pi^{n} \prod\limits_{i=1}^{n} \sqrt{1-\lambda_{i}^{2}}} \exp\left(-\xxi^{T} \left[
                                                                \begin{array}{cc}
                                                                  \left(\mathbf{I}_{n}- \mathbf{\Lambda}_{\mathbf{x}}^{2}\right)^{-1}& \mathbf{0} \\
                                                                  \mathbf{0} & \left(\mathbf{I}_{n}- \mathbf{\Lambda}_{\mathbf{x}}^{2}\right)^{-1} \\
                                                                \end{array}
                                                              \right] \xxi  \right) \times \\\nonumber
    & \hspace*{30mm} \times I_{0}\left(\left(\xxi^{T} \left[
                                                                \begin{array}{cc}
                                                                  \mathbf{D}_{\mathbf{x}} &  \mathbf{0} \\
                                                                  \mathbf{0} &  -\mathbf{D}_{\mathbf{x}}\\
                                                                \end{array}
                                                              \right]     \xxi  \right)^{2}+\left(\xxi^{T} \left[
                                                                \begin{array}{cc}
                                                                  \mathbf{0} &  \mathbf{D}_{\mathbf{x}} \\
                                                                  \mathbf{D}_{\mathbf{x}} &  \mathbf{0}\\
                                                                \end{array}
                                                              \right]     \xxi  \right)^{2}\right)  \quad \lambda_{2n}\text{-a.e.},
\end{align}
where $I_{0}(x)=\int_{0}^{1}\exp\left(x \cos (2 \pi \theta)\right) d\theta $
is the modified Bessel function of the first kind of order zero \cite{Abr65}.
Here, we have used the identity
\begin{align*}
  \int_{0}^{1}\exp\left(a \cos (2 \pi \theta) + b \sin (2 \pi \theta)\right) d\theta &=
  \int_{0}^{1}\exp\left(r \cos (2 \pi \theta_0)\cos (2 \pi \theta) +  r \sin (2 \pi \theta_0)\sin (2 \pi \theta)\right) d\theta\\
  &=\int_{0}^{1}\exp\left(r \cos (2 \pi (\theta-\theta_0) \right) d\theta\\
  &=I_{0}\left(\sqrt{a^2+b^2}\right),
\end{align*}
where $(r,\theta_0)$ denotes the polar coordinates of the complex number
$(a\! + \!\jmath b)$. According to the assumptions of the theorem and to
Theorem \ref{a1_covmat}, $\mathbf{x}_{\text{(a)}}$ is Gaussian distributed
with covariance matrix $\mathbf{C}_{\mathbf{x}_{\text{(a)}}}=\mathbf{I}_{n}$
and vanishing mean vector and complementary covariance matrix. Hence,
\eqref{a1_pdf_gauss} implies $\mathbf{D}_{\mathbf{x}}=0$ and, therefore, the
the statement.

For the general case, apply the Takagi factorization to
$\mathbf{B}_{\mathbf{x}}^{-1}\mathbf{P}_{\mathbf{x}}\mathbf{B}_{\mathbf{x}}^{-T}$,
i.e.,
$\mathbf{B}_{\mathbf{x}}^{-1}\mathbf{P}_{\mathbf{x}}\mathbf{B}_{\mathbf{x}}^{-T}
= \mathbf{Q}_{\mathbf{x}} \mathbf{\Lambda}_{\mathbf{x}}
\mathbf{Q}_{\mathbf{x}}^{T}$, and consider the random vector $\mathbf{y}
\triangleq
\mathbf{Q}_{\mathbf{x}}^{-1}\mathbf{B}_{\mathbf{x}}^{-1}\mathbf{x}$. Clearly,
$\mathbf{C}_{\mathbf{y}}=\mathbf{I}_{n}$ and
$\mathbf{P}_{\mathbf{y}}=\mathbf{\Lambda}_{\mathbf{x}}$, and both
$\mathbf{y}$ and $\mathbf{y}_{\text{(a)}}$ are Gaussian distributed. From the
special case, $\mathbf{P}_{\mathbf{y}}=\mathbf{0}$, and, in turn,
$\mathbf{P}_{\mathbf{x}}= \mathbf{B}_{\mathbf{x}}
\mathbf{Q}_{\mathbf{x}}\mathbf{P}_{\mathbf{y}}
\mathbf{Q}_{\mathbf{x}}^{T}\mathbf{B}_{\mathbf{x}}^{T}= \mathbf{0}$. \eproof

\section{Differential Entropy of Complex-Valued Random Vectors}\label{sec:entropy}
As outlined in the introduction, we are interested in bounds on the
differential entropy of complex-valued random vectors. We start with a series
of definitions, which are required for the further development of the paper.
Again, we make use of the convention $0 \log 0 \triangleq 0$ and $0 \log
\frac{0}{0} \triangleq 0$.
\begin{definition}
    The \emph{differential entropy} \emph{$h(\mathbf{x})$} of a complex-valued random vector \emph{$\mathbf{x} \in \C^{n}$}
    is defined as the differential entropy of its real representation \emph{$\mathbf{x}^{\text{(r)}}$}, i.e.,
    \emph{\begin{align*}
        h(\mathbf{x}) \triangleq  h\big(\mathbf{x}^{\text{(r)}}\big) \triangleq -\int\limits_{\R^{2n}} f_{\mathbf{x}^{\text{(r)}}}(\xxi) \log
        f_{\mathbf{x}^{\text{(r)}}}(\xxi) d\xxi,
    \end{align*}}
    provided that the integrand is integrable \emph{\cite{Hal74}}.
\end{definition}

\begin{definition}\label{def:cond_entr}
    The \emph{conditional differential entropy} \emph{$h(\mathbf{x}|\mathbf{y})$} of a complex-valued random vector \emph{$\mathbf{x} \in \C^{n}$}
    given
    a complex valued random vector \emph{$\mathbf{y} \in \C^{m}$}
    is defined as the conditional differential entropy of the real representation \emph{$\mathbf{x}^{\text{(r)}}$} given the
    real representation \emph{$\mathbf{y}^{\text{(r)}}$}, i.e.,
    \emph{\begin{align*}
        h(\mathbf{x}|\mathbf{y}) \triangleq  h\big(\mathbf{x}^{\text{(r)}}|\mathbf{y}^{\text{(r)}} \big) \triangleq -\hspace*{-3mm}\int\limits_{\R^{2n+2m}} \hspace*{-3mm}f_{\mathbf{x}^{\text{(r)}};\mathbf{y}^{\text{(r)}}}(\xxi,\eeta)
        \log \frac{f_{\mathbf{x}^{\text{(r)}};\mathbf{y}^{\text{(r)}}}(\xxi,\eeta)}{f_{\mathbf{y}^{\text{(r)}}}(\eeta)} d\xxi d\eeta,
    \end{align*}}
    provided that the integrand is integrable. Here,
    \emph{$f_{\mathbf{x}^{\text{(r)}};\mathbf{y}^{\text{(r)}}}(\xxi,\eeta)$} denotes
    the joint pdf of \emph{$\mathbf{x}^{\text{(r)}}$} and
    \emph{$\mathbf{y}^{\text{(r)}}$}, whereas
    \emph{$f_{\mathbf{y}^{\text{(r)}}}(\eeta)$}
    denotes the marginal pdf of \emph{$\mathbf{y}^{\text{(r)}}$}.
\end{definition}

\begin{definition}
    The \emph{mutual information} \emph{$I(\mathbf{x};\mathbf{y})$} between the complex-valued random vectors \emph{$\mathbf{x} \in \C^{n}$} and \emph{$\mathbf{y} \in \C^{m}$}
    is defined as the mutual information between their real representations \emph{$\mathbf{x}^{\text{(r)}}$} and \emph{$\mathbf{y}^{\text{(r)}}$}, i.e.,
    \emph{\begin{align*}
        I(\mathbf{x};\mathbf{y}) \triangleq  I\big(\mathbf{x}^{\text{(r)}};\mathbf{y}^{\text{(r)}} \big) \triangleq \hspace*{-3mm} \int\limits_{\R^{2n+2m}} \hspace*{-3mm} f_{\mathbf{x}^{\text{(r)}};\mathbf{y}^{\text{(r)}}}(\xxi,\eeta)
        \log \frac{f_{\mathbf{x}^{\text{(r)}};\mathbf{y}^{\text{(r)}}}(\xxi,\eeta)}{f_{\mathbf{x}^{\text{(r)}}}(\xxi) f_{\mathbf{y}^{\text{(r)}}}(\eeta)} d\xxi d\eeta,
    \end{align*}}
    where
    \emph{$f_{\mathbf{x}^{\text{(r)}};\mathbf{y}^{\text{(r)}}}(\xxi,\eeta)$} denotes
    the joint pdf of \emph{$\mathbf{x}^{\text{(r)}}$} and
    \emph{$\mathbf{y}^{\text{(r)}}$}, and
    \emph{$f_{\mathbf{x}^{\text{(r)}}}(\xxi)$} and \emph{$f_{\mathbf{y}^{\text{(r)}}}(\eeta)$}
    are the marginal pdfs of \emph{$\mathbf{x}^{\text{(r)}}$} and \emph{$\mathbf{y}^{\text{(r)}}$}, respectively.
\end{definition}
It is well known that these quantities satisfy the following relations,
\begin{subequations}
\begin{align}\label{mutual_inf_equ}
   I(\mathbf{x};\mathbf{y}) &= h(\mathbf{x}) - h(\mathbf{x}|\mathbf{y}) = h(\mathbf{y}) - h(\mathbf{y}|\mathbf{x}) = h(\mathbf{x})+h(\mathbf{y}) - h(\mathbf{x},\mathbf{y}),\\\label{mutual_inf_inequ}
   I(\mathbf{x};\mathbf{y}) &\geq 0,
\end{align}
\end{subequations}
with equality in \eqref{mutual_inf_inequ} if and only if $\mathbf{x}$ and
$\mathbf{y}$ are statistically independent. Furthermore, according to the
next theorem, Gaussian distributed proper random vectors are known to be
entropy maximizers.
\begin{theorem}\label{max_entropy_th_neeser}\emph{[Neeser \& Massey]} Suppose \emph{$\mathbf{x} \in \C^{n}$} is a zero-mean
complex-valued random vector with non-singular covariance matrix
\emph{$\mathbf{C}_{\mathbf{x}}$}. Then, the differential entropy of
\emph{$\mathbf{x}$} satisfies \emph{\begin{align}\label{max_entropy_th_inequ}
  h(\mathbf{x}) \leq \log \det (\pi e \mathbf{C}_{\mathbf{x}}),
\end{align}}
with equality if and only if \emph{$\mathbf{x}$} is Gaussian distributed and
circular/proper.
\end{theorem}
\bproof See e.g., \cite{neeser93,Telatar95}. \eproof {\bf Remarks.} Let us
assume, for the moment, that $\mathbf{x}$ is known to be non-Gaussian.
Clearly, the inequality \eqref{max_entropy_th_inequ} is strict in this case
and $\log \det (\pi e \mathbf{C}_{\mathbf{x}})$ is not a tight upper bound
for the differential entropy $h(\mathbf{x})$. Similarly, if $\mathbf{x}$ is
known to be improper, the differential entropy $h(\mathbf{x})$ is strictly
smaller than $\log \det (\pi e \mathbf{C}_{\mathbf{x}})$. Loosely speaking,
there are two sources that decrease the differential entropy of a
complex-valued random vector: non-Gaussianity and improperness. In the
following, we will derive improved maximum entropy theorems that take this
observation into account. While their application is not limited to the
non-Gaussian and improper case, the obtained upper bounds are in general
tighter for these two scenarios than the upper bound given by Theorem
\ref{max_entropy_th_neeser}.

\subsection{Maximum Entropy Theorem I}
We first prove a maximum entropy theorem that is especially suited to the
non-Gaussian case. However, also for Gaussian distributed random vectors the
obtained upper bound will turn out to be tighter than the one of Theorem
\ref{max_entropy_th_neeser}. It associates a specific circular random vector
to a given random vector and upper bounds the differential entropy of the
given random vector  by the differential entropy of the associated circular
random vector. 
\begin{theorem}\label{max_entropy_th_taubI}\emph{(Maximum Entropy Theorem for Complex-Valued Random
Vectors I)} Suppose \emph{$\mathbf{x} \in \C^{n}$} is a complex-valued random
vector. Then, the differential entropies of \emph{$\mathbf{x}$} and its
circular analog \emph{$\mathbf{x}_{\text{(a)}}$} satisfy \emph{\begin{align*}
  h(\mathbf{x}) \leq h\left(\mathbf{x}_{\text{(a)}}\right),
\end{align*}}
with equality if and only if \emph{$\mathbf{x}$} is circular.
\end{theorem}
\bproof Since $\mathbf{x}_{\text{(a)}}= e^{\jmath 2 \pi \psi} \mathbf{x}$
with $\psi$ independent of $\mathbf{x}$, we have for fixed (deterministic)
$\varphi$,
\begin{align*}
  h(\mathbf{x}_{\text{(a)}}| \psi=\varphi)=h(e^{\jmath 2 \pi \varphi}
\mathbf{x})=h(\mathbf{x}),
\end{align*}
and, furthermore, by applying Fubini's theorem to Definition
\ref{def:cond_entr},
\begin{align*}
  h(\mathbf{x}_{\text{(a)}}| \psi)=\int h(\mathbf{x}_{\text{(a)}}| \psi=\varphi) f_{\psi}(\varphi) d\varphi  =\int
h(\mathbf{x})f_{\psi}(\varphi) d\varphi = h(\mathbf{x}).
\end{align*}
Therefore,
\begin{align}\label{entr_diff}
  h(\mathbf{x}_{\text{(a)}})-h(\mathbf{x}) = h(\mathbf{x}_{\text{(a)}})-h(\mathbf{x}_{\text{(a)}}| \psi) = I(\mathbf{x}_{\text{(a)}}; \psi) \geq 0,
\end{align}
where we have used \eqref{mutual_inf_equ} and \eqref{mutual_inf_inequ}.
$\mathbf{x}_{\text{(a)}}$ and $\psi$ are independent, i.e.,
$h(\mathbf{x}_{\text{(a)}})=h(\mathbf{x})$, if and only if
$\mathbf{x}_{\text{(a)}}^{\text{(s)}}$ and $\psi$ are independent. To
investigate this independence,\footnote{The following technical derivation is
required in order to show identical distributions of $e^{\jmath 2 \pi
\theta}\mathbf{x}$ for all $\theta \in [0,1[$ and not only for
$\theta$\hspace*{1mm}$\lambda_{1}\text{-a.e.}$ on $[0,1[$.} consider the
joint pdf of $\mathbf{x}_{\text{(a)}}^{\text{(s)}}$ and $\psi$, i.e.,
\begin{align*}
  f_{\mathbf{x}^{\text{(s)}}_{\text{(a)}};\psi}(r_1,\ldots,r_n,\phi_1,\ldots,\phi_n,\varphi)
&=f_{\mathbf{x}^{\text{(s)}}_{\text{(a)}}|\psi}(r_1,\ldots\ldots,r_n,\phi_1,\ldots,\phi_n|\varphi)\,f_{\psi}(\varphi)\\
&=f_{\mathbf{x}^{\text{(s)}}}\left(r_1,\ldots,r_n,\phi_1,\ldots,\phi_{n-1},[\phi_n-\varphi]_{[0,1[}\right)\\
&=f_{\tilde{\mathbf{x}}^{\text{(s)}}}(\tilde{\xxi}) f_{\vartheta|\tilde{\mathbf{x}}^{\text{(s)}}}\left([\phi_n-\varphi]_{[0,1[}\big|\tilde{\xxi}\right),
\end{align*}
where $\tilde{\xxi}\triangleq(r_1,\ldots,r_n,\phi_1,\ldots,\phi_{n-1}) \in
\widetilde{\mathcal{S}}_{\mathbf{x}}$, $\widetilde{\mathcal{S}}_{\mathbf{x}}
\subset (\R_{0}^{+})^{n}\times ([0,1[)^{n-1}$ being the support set of
$f_{\tilde{\mathbf{x}}^{\text{(s)}}}$, cf. \eqref{cond_dens}. Since
$f_{\mathbf{x}^{\text{(s)}}_{\text{(a)}}}(\tilde{\xxi},\phi_n)=f_{\tilde{\mathbf{x}}^{\text{(s)}}}(\tilde{\xxi})$\hspace*{1mm}$\lambda_{2n}\text{-a.e.}$
on $\widetilde{\mathcal{S}}_{\mathbf{x}}\times  [0,1[$ and
$f_{\psi}(\varphi)=1$\hspace*{1mm}$\lambda_{1}\text{-a.e.}$ on $[0,1[$,
independence of $\mathbf{x}_{\text{(a)}}^{\text{(s)}}$ and $\psi$ is
equivalent to
\begin{align}\label{indep_cond}
    f_{\vartheta|\tilde{\mathbf{x}}^{\text{(s)}}}\left([\phi_n-\varphi]_{[0,1[}\big|\tilde{\xxi}\right) &= 1 \quad \lambda_{2n+1}\text{-a.e.  on
    $[0,1[^{2}\times \widetilde{\mathcal{S}}_{\mathbf{x}}$ as function of
    $(\phi_n,\varphi,\tilde{\xxi})$}.
\end{align}
Transforming both sides of this equation according to $\phi'_{n}\triangleq
[\phi_n-\varphi]_{[0,1[}$ and $\varphi' \triangleq \varphi$, a similar
partitioning argument as in the proof of Lemma \ref{lem:sheared_distr} shows
that \eqref{indep_cond} is equivalent to
\begin{align}\label{indep_cond1}
    f_{\vartheta|\tilde{\mathbf{x}}^{\text{(s)}}}\left(\phi'_n\big|\tilde{\xxi}\right) = 1 \quad \lambda_{2n+1}\text{-a.e.  on
$[0,1[^{2}\times \widetilde{\mathcal{S}}_{\mathbf{x}}$ as function of
$(\phi'_n,\varphi',\tilde{\xxi})$}.
\end{align}
Marginalization of both sides of \eqref{indep_cond1} with respect to
$\varphi'$ yields
\begin{align*}
    f_{\vartheta|\tilde{\mathbf{x}}^{\text{(s)}}}\left(\phi'_n\big|\tilde{\xxi}\right) = 1 \quad \lambda_{2n}\text{-a.e.  on
$[0,1[\times \widetilde{\mathcal{S}}_{\mathbf{x}}$ as function of
$(\phi'_n,\tilde{\xxi})$},
\end{align*}
so that---according to Corollary \ref{cor:circ_sheared} and
\eqref{cond_dens}---equality $h(\mathbf{x}_{\text{(a)}})=h(\mathbf{x})$
implies circularity of $\mathbf{x}$. The converse statement follows from
Theorem \ref{a1_charac_div}.\eproof {\bf Remarks.} Since
$\mathbf{x}_{\text{(a)}}$ is non-Gaussian in general,\footnote{Note that it
is possible to define an improper (non-Gaussian) random vector, such that its
circular analog is Gaussian distributed. In this case, Theorem
\ref{max_entropy_th_taubI} does not yield an improvement over Theorem
\ref{max_entropy_th_neeser}.} the upper bound in Theorem
\ref{max_entropy_th_taubI} is typically tighter than the upper bound in
Theorem \ref{max_entropy_th_neeser}. Furthermore, Theorem
\ref{max_entropy_th_taubI} does not need the requirement of finite
second-order moments. The next corollary states that for improper Gaussian
distributed random vectors the upper bound in Theorem
\ref{max_entropy_th_taubI} is strictly smaller than the upper bound in
Theorem \ref{max_entropy_th_neeser}.
\begin{corollary}\label{cor:comp_bounds}
  Suppose \emph{$\mathbf{x} \in \C^{n}$} is a zero-mean,
complex-valued, and Gaussian distributed random vector with non-singular
covariance matrix \emph{$\mathbf{C}_{\mathbf{x}}$}, such that
\emph{$\|\mathbf{B}_{\mathbf{x}}^{-1}\mathbf{P}_{\mathbf{x}}\mathbf{B}_{\mathbf{x}}^{-T}\|_{2}<1$},
where $\emph{$\mathbf{B}_{\mathbf{x}}$}$ denotes a generalized Cholesky
factor of \emph{$\mathbf{C}_{\mathbf{x}}$} and
\emph{$\mathbf{P}_{\mathbf{x}}$} denotes the complementary covariance matrix
of \emph{$\mathbf{x}$}. Then, the differential entropy of its circular analog
 \emph{$\mathbf{x}_{\text{(a)}}$} satisfies
\emph{\begin{align*}
  h\left(\mathbf{x}_{\text{(a)}}\right) \leq \log \det (\pi e
  \mathbf{C}_{\mathbf{x}}),
\end{align*}}
with equality if and only if \emph{$\mathbf{x}$} is proper.
\end{corollary}
\bproof Since $\mathbf{x}_{\text{(a)}}$ is zero-mean with covariance matrix
$\mathbf{C}_{\mathbf{x}_{\text{(a)}}}=\mathbf{C}_{\mathbf{x}}$, cf. Theorem
\ref{a1_covmat}, the inequality follows from Theorem
\ref{max_entropy_th_neeser}. Furthermore, equality
$h\left(\mathbf{x}_{\text{(a)}}\right) = \log \det (\pi e
\mathbf{C}_{\mathbf{x}})$ implies Gaussianity of $\mathbf{x}_{\text{(a)}}$,
and, according to Theorem \ref{a1_distr_gauss}, properness of $\mathbf{x}$.
\eproof

\subsection{Maximum Entropy Theorem II}
Here, we prove a maximum entropy theorem that is especially suited to the
improper case. The derivation is based on a maximum entropy theorem for
\emph{real-valued} random vectors.
\begin{theorem}\label{max_entropy_real}\emph{(Maximum Entropy Theorem for Real-Valued Random
Vectors)} Suppose \emph{$\mathbf{x} \in \R^{n}$} is a \emph{real-valued}
random vector with non-singular
     covariance matrix $\mathbf{C_{x}}$. Then, the differential entropy  of $\mathbf{x}$
     satisfies
     \begin{equation}\nonumber
         h(\mathbf{x}) \leq \frac{1}{2}\log\det\left(2 \pi e
    \mathbf{C_{x}}\right)
     \end{equation}
     with equality if and only if $\mathbf{x}$ is Gaussian distributed.
\end{theorem}
\bproof For the proof of this theorem for $\mathbf{x}$ being zero-mean see
e.g., \cite{Cover91}. The general case, where $\mathbf{x}$ has a
non-vanishing mean vector, follows immediately since both differential
entropy and covariance matrix are invariant with respect to translations.
\eproof We are now able to state the main theorem of this section.
\begin{theorem}\label{max_entropy_th_taubII}\emph{(Maximum Entropy Theorem for Complex-Valued Random
Vectors II)} Suppose \emph{$\mathbf{x} \in \C^{n}$} is a complex-valued
random vector with non-singular covariance matrix
\emph{$\mathbf{C}_{\mathbf{x}}$}, such that
\emph{$\|\mathbf{B}_{\mathbf{x}}^{-1}\mathbf{P}_{\mathbf{x}}\mathbf{B}_{\mathbf{x}}^{-T}\|_{2}<1$},
where $\emph{$\mathbf{B}_{\mathbf{x}}$}$ denotes a generalized Cholesky
factor of \emph{$\mathbf{C}_{\mathbf{x}}$} and
\emph{$\mathbf{P}_{\mathbf{x}}$} denotes the complementary covariance matrix
of \emph{$\mathbf{x}$}. Then, the differential entropy of \emph{$\mathbf{x}$}
satisfies \emph{\begin{align*}
  h(\mathbf{x}) \leq \log \det (\pi e \mathbf{C}_{\mathbf{x}})+\frac{1}{2} \sum\limits_{i=1}^{n} \log(1-\lambda_{i}^{2}),
\end{align*}}
where $\lambda_{i}$ are the singular values of
$\mathbf{B}_{\mathbf{x}}^{-1}\mathbf{P}_{\mathbf{x}}\mathbf{B}_{\mathbf{x}}^{-T}$,
with equality if and only if \emph{$\mathbf{x}$} is Gaussian distributed.
\end{theorem}
\bproof According to Theorem \ref{max_entropy_real},
\begin{eqnarray*}
  h(\mathbf{x})\hspace*{-3mm} &\leq& \hspace*{-3mm} \frac{1}{2}\log\det\left(2 \pi e
    \mathbf{C}_{\mathbf{x}^{\text{(r)}}}\right)\\
    \hspace*{-3mm}&\overset{\eqref{expr_cov_compl}}{=}&\hspace*{-3mm}\frac{1}{2}\log\det\left(\pi e
    \left(\overline{\mathbf{C}}_{\mathbf{x}} + \underline{\mathbf{P}}_{\mathbf{x}} \right)\right)\\
    \hspace*{-3mm}&=& \hspace*{-3mm} n \log(\pi e) + \frac{1}{2}\log\det\left(\overline{\mathbf{B}_{\mathbf{x}} \mathbf{B}_{\mathbf{x}}^{H}} + \underline{\mathbf{P}}_{\mathbf{x}} \right)\\
    \hspace*{-3mm}&\overset{\eqref{alg_a},\eqref{alg_c}}{=}& \hspace*{-3mm} n \log(\pi e) + \frac{1}{2}\log\det\left(\overline{\mathbf{B}}_{\mathbf{x}} \overline{\mathbf{B}}_{\mathbf{x}}^{T} + \underline{\mathbf{P}}_{\mathbf{x}} \right)\\
    \hspace*{-3mm}&\overset{\eqref{alg_e}}{=}&\hspace*{-3mm} \log\det(\pi e \mathbf{C}_{\mathbf{x}}) + \frac{1}{2}\log\det\left(\mathbf{I}_{2n}  + \overline{\mathbf{B}}_{\mathbf{x}} ^{-1}\underline{\mathbf{P}}_{\mathbf{x}} \overline{\mathbf{B}}_{\mathbf{x}}^{-T} \right)\\
    \hspace*{-3mm}&\overset{\eqref{alg_a},\eqref{alg_b},\eqref{alg_c}}{=}& \hspace*{-3mm}\log\det(\pi e \mathbf{C}_{\mathbf{x}}) + \frac{1}{2}\log\det\left(\mathbf{I}_{2n}  + \underline{\mathbf{B}_{\mathbf{x}}^{-1}\mathbf{P}_{\mathbf{x}} \mathbf{B}_{\mathbf{x}}^{-T}} \right)\\
    \hspace*{-3mm}&=&\hspace*{-3mm} \log\det(\pi e \mathbf{C}_{\mathbf{x}}) + \frac{1}{2}\log\prod\limits_{i=1}^{n} (1-\lambda_{i}^{2}),
\end{eqnarray*}
where the last identity follows from Proposition \ref{pro:eigenv_compl}
applied to the random vector $\mathbf{y} \triangleq
\mathbf{B}_{\mathbf{x}}^{-1} \mathbf{x}$. Note that
$\mathbf{P}_{\mathbf{y}}=\mathbf{B}_{\mathbf{x}}^{-1}\mathbf{P}_{\mathbf{x}}
\mathbf{B}_{\mathbf{x}}^{-T}$. We also conclude from the last expression that
the non-singularity of $\mathbf{C}_{\mathbf{x}^{\text{(r)}}}$, which is
required for the application of Theorem \ref{max_entropy_real}, is a direct
consequence of the assumption
$\|\mathbf{B}_{\mathbf{x}}^{-1}\mathbf{P}_{\mathbf{x}}\mathbf{B}_{\mathbf{x}}^{-T}\|_{2}<1$,
since $\lambda_{i}\leq
\|\mathbf{B}_{\mathbf{x}}^{-1}\mathbf{P}_{\mathbf{x}}\mathbf{B}_{\mathbf{x}}^{-T}\|_{2}$.
The equality criterion is obvious.\eproof {\bf Remarks.} Note that
$\frac{1}{2} \sum\limits_{i=1}^{n} \log(1-\lambda_{i}^{2})\leq 0$ with
equality if and only if $\mathbf{x}$ is proper, so that Theorem
\ref{max_entropy_th_taubII} implies Theorem \ref{max_entropy_th_neeser}. The
upper bound in Theorem \ref{max_entropy_th_taubII} is the differential
entropy of a Gaussian distributed but in general non-circular/improper random
vector with same covariance matrix and complementary covariance matrix as
$\mathbf{x}$, whereas the upper bound in Theorem \ref{max_entropy_th_taubI}
is the differential entropy of a circular but in general non-Gaussian random
vector with same covariance matrix as $\mathbf{x}$. Which of the two bounds
is tighter depends on the situation, i.e., on the degree of improperness and
non-Gaussianity; a general statement is not possible. However, for an
improper Gaussian distributed random vector $\mathbf{x}$,
\begin{align*}
    \log \det (\pi e \mathbf{C}_{\mathbf{x}})+\frac{1}{2} \sum\limits_{i=1}^{n} \log(1-\lambda_{i}^{2}) <  h(\mathbf{x}_{\text{(a)}}),
\end{align*}
whereas for a circular non-Gaussian random vector $\mathbf{x}$,
\begin{align*}
    h(\mathbf{x}_{\text{(a)}}) < \log \det (\pi e \mathbf{C}_{\mathbf{x}})+\frac{1}{2} \sum\limits_{i=1}^{n} \log(1-\lambda_{i}^{2}) = \log \det (\pi e \mathbf{C}_{\mathbf{x}}).
\end{align*}

\section{Capacity of Complex-Valued Channels}\label{sec:capacity}
In this section we study the influence of
circularity/properness---non-circularity/improperness on channel capacity. In
particular, we investigate vector-valued (MIMO) channels with complex-valued
input and complex-valued output. For simplicity, we only consider linear
channels with additive noise, i.e., channels of the form
\begin{align}\label{MIMO_channel_model}
  \mathbf{y}=\mathbf{H} \mathbf{x} +\mathbf{z},
\end{align}
where $\mathbf{x}\in\C^{m}$, $\mathbf{y}\in\C^{n}$, and $\mathbf{z}\in\C^{n}$
denote transmit, receive, and noise vector, respectively, and $\mathbf{H} \in
\C^{n \times m}$ is the channel matrix. Both\footnote{Without loss of
generality, since an iid (with respect to channel uses) $\mathbf{x}$ is
capacity achieving if $\mathbf{z}$ and $\mathbf{H}$ (if applicable) are iid.}
$\mathbf{x}$ and $\mathbf{z}$ are modeled as iid (only with respect to
channel uses; within the random vectors the iid assumption is not made)
vector-valued random processes, whereas $\mathbf{H}$ is either assumed to be
deterministic or is modeled as an iid (again, only with respect to channel
uses) matrix-valued random process. Furthermore, $\mathbf{x}$, $\mathbf{z}$,
and $\mathbf{H}$ (if applicable) are assumed to be statistically independent.
Note that the assumption of a Gaussian distributed noise vector $\mathbf{z}$
is only made for the special case investigated in Section
\ref{sec:improp_noise} but not in general. The channel is characterized by
the conditional distribution of $\mathbf{y}$ given $\mathbf{x}$ via the
conditional pdf
$f_{\mathbf{y}^{\text{(r)}}|\mathbf{x}^{\text{(r)}}}(\eeta|\xxi)$  of their
real representations $\mathbf{y}^{\text{(r)}}$ given
$\mathbf{x}^{\text{(r)}}$, as well as by a set $\mathcal{I}$ of admissible
input distributions. We write $\mathbf{x} \in \mathcal{I}$, if the
distribution of $\mathbf{x}$ defined by the pdf $f_{\mathbf{x}^{\text{(r)}}}$
is in $\mathcal{I}$. Then, the \emph{capacity/noncoherent capacity} of
\eqref{MIMO_channel_model} is given by the supremum of the mutual information
over the set of admissible input distributions \cite{BobGray10}, i.e., by
\begin{align*}
  C=\sup_{\mathbf{x} \in \mathcal{I}}I(\mathbf{x};\mathbf{y}).
\end{align*}
If, for the case of a random channel matrix, it is additionally assumed that
the channel realizations are known to the receiver (but not to the
transmitter), the channel output of \eqref{MIMO_channel_model} is the pair
\begin{align}\label{MIMO_channel_model_coherent}
  (\mathbf{y},\mathbf{H})=(\mathbf{H} \mathbf{x} +\mathbf{z},\mathbf{H}),
\end{align}
so that the channel law of \eqref{MIMO_channel_model_coherent} is governed by
the conditional pdf
$f_{\mathbf{y}^{\text{(r)}};\mathbf{H}^{\text{(r)}}|\mathbf{x}^{\text{(r)}}}(\eeta,\cchi|\xxi)$,
where $\mathbf{H}^{\text{(r)}}$ is defined by an appropriate stacking of real
and imaginary part of $\mathbf{H}$. Therefore, the \emph{coherent capacity}
of \eqref{MIMO_channel_model_coherent} is given by
\begin{align*}
  C_{\text{c}}=\sup_{\mathbf{x} \in \mathcal{I}}I(\mathbf{x};\mathbf{y},\mathbf{H})=
  \sup_{\mathbf{x} \in \mathcal{I}} \int I(\mathbf{x}^{\text{(r)}};\mathbf{y^{\text{(r)}}}\big|\mathbf{H}^{\text{(r)}}=\cchi) f_{\mathbf{H}^{\text{(r)}}}(\cchi) d\cchi,
\end{align*}
where $f_{\mathbf{H}^{\text{(r)}}}(\cchi)$ denotes the pdf of
$\mathbf{H}^{\text{(r)}}$ and Fubini's Theorem has been used. A random vector
$\mathbf{x} \in \mathcal{I}$ is said to be \emph{capacity-achieving} for
\eqref{MIMO_channel_model} or \eqref{MIMO_channel_model_coherent}, if
$I(\mathbf{x};\mathbf{y})=C$ or
$I(\mathbf{x};\mathbf{y},\mathbf{H})=C_{\text{c}}$, respectively.

\subsection{Circular Noise Vector}\label{sec:circ_noise}
Here, we assume that the noise vector $\mathbf{z}\in\C^{n}$ is circular and
that $\mathcal{I}$ is closed under the operation of forming the circular
analog, i.e., that $\mathbf{x} \in \mathcal{I}$ implies
$\mathbf{x}_{\text{(a)}} \in \mathcal{I}$---in the following shortly termed
\emph{circular-closed}. Note that this closeness assumption is a natural
assumption, since the operation of forming the circular analog of the first
kind preserves both peek and average power constraints, cf. Theorem
\ref{a1_covmat}, which are the most common constraints for defining
$\mathcal{I}$.  If $\mathbf{z}$ is Gaussian distributed, it has been shown in
\cite{Telatar95} that capacity (for deterministic $\mathbf{H}$) and coherent
capacity (for random $\mathbf{H}$) are achieved by circular (Gaussian
distributed) random vectors, respectively. The proofs are based on Theorem
\ref{max_entropy_th_neeser}. The following Theorems \ref{Th:cap1} and
\ref{Th:cap3} extend these results to the non-Gaussian case.
\begin{theorem}\label{Th:cap1}
  Suppose for \eqref{MIMO_channel_model} a deterministic channel matrix \emph{$\mathbf{H}\in
\C^{n \times m}$}, a circular noise vector
  \emph{$\mathbf{z}\in\C^{n}$}, and a circular-closed set \emph{$\mathcal{I}$} of admissible input
  distributions. Then, there exists a circular random
  vector $\mathbf{x}\in\C^{m}$ that achieves the capacity of \eqref{MIMO_channel_model}.
\end{theorem}\bproof Let us denote by $\mathbf{x}'\in\mathcal{I}\, $ a---not necessarily circular---capacity-achieving random
vector. 
According to \eqref{mutual_inf_equ}, its circular analog
$\mathbf{x}\triangleq\mathbf{x}'_{\text{(a)}}= e^{\jmath 2
    \pi \psi}  \mathbf{x}' \in\mathcal{I}$, where $\psi \in [0,1[$ is
    uniformly distributed and assumed to be
    independent of
    $\mathbf{x}'$ and $\mathbf{z}$, satisfies
\begin{align*}
  I(\mathbf{x};\mathbf{y}) &= h(\mathbf{y}) - h\left(\mathbf{y}|\mathbf{x}\right)\\
    &= h\left(\mathbf{H}\mathbf{x} +\mathbf{z}\right) - h\left(\mathbf{H}\mathbf{x} +\mathbf{z}|\mathbf{x}\right)\\
    &= h\left(\mathbf{H}e^{\jmath 2
    \pi \psi}  \mathbf{x}' +\mathbf{z}\right) - h\left(\mathbf{z}\right)\\
    &= h\left(e^{\jmath 2
    \pi \psi} (\mathbf{H} \mathbf{x}' + e^{-\jmath 2
    \pi \psi} \mathbf{z})\right) - h\left(\mathbf{z}\right).
\end{align*}
Note that $\mathbf{z}_{\text{(a)}}= e^{-\jmath 2 \pi \psi} \mathbf{z} =
\mathbf{z}$, cf. Theorem \ref{a1_charac_div}, and that
$\mathbf{z}_{\text{(a)}}$ is independent of $\psi$, according to
\eqref{entr_diff}. Therefore,
\begin{align*}
  I(\mathbf{x};\mathbf{y})
    &= h\left(\left(\mathbf{H} \mathbf{x}' +  \mathbf{z}\right)_{\text{(a)}}\right) - h\left(\mathbf{z}\right)\\
    &\overset{(*)}{\geq} h\left(\mathbf{H} \mathbf{x}' +  \mathbf{z}\right) - h\left(\mathbf{z}\right)\\
    &=I(\mathbf{x}';\mathbf{H}\mathbf{x}' +\mathbf{z})\\
    &=C,
\end{align*}
where $(*)$ follows from Theorem \ref{max_entropy_th_taubI}. Hence, the
circular $\mathbf{x}$ is capacity-achieving.\eproof
\begin{theorem}\label{Th:cap2}
  Suppose for \eqref{MIMO_channel_model} a random channel matrix \emph{$\mathbf{H}\in
\C^{n \times m}$}, a circular noise vector
  \emph{$\mathbf{z}\in\C^{n}$}, and a circular-closed set \emph{$\mathcal{I}$} of admissible input
  distributions. Then, there exists a circular random
  vector $\mathbf{x}\in\C^{m}$ that achieves the noncoherent capacity of \eqref{MIMO_channel_model}.
\end{theorem}\bproof Let us denote by $\mathbf{x}'\in\mathcal{I}\, $ a---not necessarily circular---capacity-achieving random
vector and let $\mathbf{ x}_{(\theta)} \triangleq e^{\jmath 2 \pi
\theta}\mathbf{x}'$ (with $\theta\in [0,1[$ being deterministic). With
$\mathbf{y}_{(\theta)} =\mathbf{H} \mathbf{x}_{(\theta)}  +\mathbf{z}$ we
obtain,
\begin{align*}
  I\left(\mathbf{x}_{(\theta)};\mathbf{y}_{(\theta)}\right) &= h\left(\mathbf{y}_{(\theta)}\right) - h\left(\mathbf{y}_{(\theta)}\big|\mathbf{x}_{(\theta)}\right)\\
    &= h\left(\mathbf{H}\mathbf{x}_{(\theta)} +\mathbf{z}\right) - h\left(\mathbf{H}\mathbf{x}_{(\theta)} +\mathbf{z}\big|\mathbf{x}_{(\theta)}\right)\\
    &\overset{(*)}{=}h\left(e^{\jmath 2 \pi \theta}(\mathbf{H}\mathbf{x}' +e^{-\jmath 2 \pi
\theta}\mathbf{z})\right) - \int h\left(\mathbf{H}\mathbf{x}_{(\theta)} +\mathbf{z}\big|\mathbf{x}_{(\theta)}^{\text{(r)}} = \xxi \right)f_{\mathbf{x}_{(\theta)}^{\text{(r)}}}(\xxi) d\xxi\\
&= h\left(e^{\jmath 2 \pi \theta}(\mathbf{H}\mathbf{x}' +e^{-\jmath 2 \pi
\theta}\mathbf{z})\right) - \int h\left(\mathbf{H}e^{\jmath 2 \pi
\theta}\mathbf{x}' +\mathbf{z}\big|{\mathbf{x}' }^{\text{(r)}} = \xxi \right)f_{{\mathbf{x}' }^{\text{(r)}}}(\xxi) d\xxi\\
&= h\left(e^{\jmath 2 \pi \theta}(\mathbf{H}\mathbf{x}' +e^{-\jmath 2 \pi
\theta}\mathbf{z})\right) - \int h\left(e^{\jmath 2 \pi
\theta}(\mathbf{H}\mathbf{x}' +e^{-\jmath 2 \pi
\theta}\mathbf{z})\big|{\mathbf{x}' }^{\text{(r)}} = \xxi \right)f_{{\mathbf{x}' }^{\text{(r)}}}(\xxi) d\xxi,
\end{align*}
where $(*)$ follows from Fubini's Theorem. Since the differential entropy of
a complex-valued random vector is invariant with respect to a multiplication
with $e^{\jmath 2 \pi \theta}$,
\begin{align*}
  I\left(\mathbf{x}_{(\theta)};\mathbf{y}_{(\theta)}\right)
&= h\left(\mathbf{H}\mathbf{x}' +e^{-\jmath 2 \pi
\theta}\mathbf{z}\right) - \int h\left(\mathbf{H}\mathbf{x}' +e^{-\jmath 2 \pi
\theta}\mathbf{z}\big|{\mathbf{x}' }^{\text{(r)}} = \xxi \right)f_{{\mathbf{x}' }^{\text{(r)}}}(\xxi) d\xxi\\
&\overset{(*)}{=} h\left(\mathbf{H}\mathbf{x}' +\mathbf{z}\right) - \int h\left(\mathbf{H}\mathbf{x}' +
\mathbf{z}\big|{\mathbf{x}' }^{\text{(r)}} = \xxi \right)f_{{\mathbf{x}' }^{\text{(r)}}}(\xxi) d\xxi\\
&=I\left(\mathbf{x}';\mathbf{y}'\right),
\end{align*}
where $\mathbf{y}' =\mathbf{H} \mathbf{x}'  +\mathbf{z}$ and $(*)$ follows
from the circularity of $\mathbf{z}$. Hence, $\mathbf{ x}_{(\theta)}$ is
capacity-achieving. It is well known that the mutual information is a concave
function with respect to the input distribution for fixed channel law
\cite{BobGray10}. Therefore, by Jensen's inequality \cite{Ash00}, the random
vector $\mathbf{x}\in\C^{m}$ with distribution defined according to
$f_{\mathbf{x}^{\text{(r)}}}(\xxi)\triangleq\int_0^1
f_{\mathbf{x}_{(\theta)}^{\text{(r)}}}(\xxi)
d\theta$\hspace*{1mm}$\lambda_{2n}\text{-a.e.}$ satisfies
\begin{align*}
  I\left(\mathbf{x};\mathbf{y}\right) \geq \int_0^1 I\left(\mathbf{x}_{(\theta)};\mathbf{y}_{(\theta)}\right) d\theta =
  \int_0^1 I\left(\mathbf{x}';\mathbf{y}'\right) d\theta = I\left(\mathbf{x}';\mathbf{y}'\right),
\end{align*}
so that $\mathbf{x}$ achieves the noncoherent capacity of
\eqref{MIMO_channel_model}. But $f_{\mathbf{x}_{(\theta)}^{\text{(r)}}}(\xxi)
=f_{{\mathbf{x}'}_{\text{(a)}}^{\text{(r)}}|\psi}(\xxi|\theta)$\hspace*{1mm}$\lambda_{2n}\text{-a.e.}$,
where $\psi$ denotes the uniformly distributed random variable used for
defining ${\mathbf{x}'}_{\text{(a)}}$ (see Definition \ref{def:a1}), and,
therefore, $\mathbf{x}={\mathbf{x}'}_{\text{(a)}} \in \mathcal{I}$. \eproof
\begin{theorem}\label{Th:cap3}
  Suppose for \eqref{MIMO_channel_model} a random channel matrix \emph{$\mathbf{H}\in
\C^{n \times m}$}, a circular noise vector
  \emph{$\mathbf{z}\in\C^{n}$}, and a circular-closed set \emph{$\mathcal{I}$} of admissible input
  distributions. Then, there exists a circular random
  vector $\mathbf{x}\in\C^{m}$ that achieves the coherent capacity of \eqref{MIMO_channel_model_coherent}.
\end{theorem}\bproof Let us denote by $\mathbf{x}'\in\mathcal{I}\, $ a---not necessarily circular---random
vector that achieves the coherent capacity of \eqref{MIMO_channel_model_coherent}. 
Using the same line of arguments as in the proof of Theorem \ref{Th:cap1},
its circular analog $\mathbf{x}\triangleq\mathbf{x}'_{\text{(a)}}= e^{\jmath
2
    \pi \psi}  \mathbf{x}' \in\mathcal{I}$, where $\psi \in [0,1[$ is
    uniformly distributed and assumed to be
    independent of
    $\mathbf{x}'$, $\mathbf{z}$, and $\mathbf{H}$, can be shown to satisfy
\begin{align*}
  I(\mathbf{x}^{\text{(r)}};\mathbf{y^{\text{(r)}}}\big|\mathbf{H}^{\text{(r)}}=\cchi) \geq I({\mathbf{x}'}^{\text{(r)}};\mathbf{{y'}^{\text{(r)}}}\big|\mathbf{H}^{\text{(r)}}=\cchi),
\end{align*}
where $\mathbf{y}=\mathbf{H} \mathbf{x} +\mathbf{z}$ and
$\mathbf{y}'=\mathbf{H} \mathbf{x}' +\mathbf{z}$. It follows that
\begin{align*}
  I(\mathbf{x};\mathbf{y},\mathbf{H})&=
  \int I(\mathbf{x}^{\text{(r)}};\mathbf{y^{\text{(r)}}}\big|\mathbf{H}^{\text{(r)}}=\cchi) f_{\mathbf{H}^{\text{(r)}}}(\cchi) d\cchi\\ &\geq \int I({\mathbf{x}'}^{\text{(r)}};\mathbf{{y'}^{\text{(r)}}}\big|\mathbf{H}^{\text{(r)}}=\cchi) f_{\mathbf{H}^{\text{(r)}}}(\cchi) d\cchi\\
  &= I(\mathbf{x}';\mathbf{y}',\mathbf{H})\\
  &=C_{\text{c}},
\end{align*}
i.e., $\mathbf{x}$ achieves the coherent capacity of
\eqref{MIMO_channel_model_coherent}.\eproof

\subsection{Circular Channel Matrix}
Here, we assume that the channel matrix $\mathbf{H} \in \C^{n \times m}$ is
random, and---additionally---that an arbitrary stacking of the elements of
$\mathbf{H}$ into an $n m$-dimensional vector yields a circular random
vector. The noise vector $\mathbf{z}$ is not required to be circular. Note
that this is the opposite situation compared with Section
\ref{sec:circ_noise}, where $\mathbf{z}$ is circular but $\mathbf{H}$ is
arbitrary.  Again, it is assumed that the set $\mathcal{I}$ of admissible
input distributions is circular-closed. For the input distributions that
achieve the noncoherent capacity of \eqref{MIMO_channel_model} and the
coherent capacity of \eqref{MIMO_channel_model_coherent}, respectively, we
have the following results.
\begin{theorem}\label{Th:cap4}
  Suppose for \eqref{MIMO_channel_model} a random channel matrix \emph{$\mathbf{H}\in
\C^{n \times m}$}, such that the random vector, which is obtained from an
arbitrary stacking of the elements of $\mathbf{H}$ into an $n m$-dimensional
vector, is circular, and a circular-closed set \emph{$\mathcal{I}$} of
admissible input
  distributions. Then, there exists a circular random
  vector $\mathbf{x}\in\C^{m}$ that achieves the noncoherent capacity of \eqref{MIMO_channel_model}.
\end{theorem}\bproof Let us denote by $\mathbf{x}'\in\mathcal{I}\, $ a---not necessarily circular---random
vector that achieves the noncoherent capacity of \eqref{MIMO_channel_model},
and let $\mathbf{x}\triangleq\mathbf{x}'_{\text{(a)}}= e^{\jmath 2 \pi \psi}
\mathbf{x}' \in\mathcal{I}$, where $\psi \in [0,1[$ is uniformly distributed
and assumed to be independent of $\mathbf{x}'$, $\mathbf{z}$, and
$\mathbf{H}$, be its circular analog. We have,
\begin{align*}
  I(\mathbf{x};\mathbf{y}) &=h\left(\mathbf{H}\mathbf{x} +\mathbf{z}\right) - h\left(\mathbf{H}\mathbf{x} +\mathbf{z}|\mathbf{x}\right),\\
  I(\mathbf{x}';\mathbf{y}') &=h\left(\mathbf{H}\mathbf{x}' +\mathbf{z}\right) - h\left(\mathbf{H}\mathbf{x}' +\mathbf{z}|\mathbf{x}'\right),
\end{align*}
and intend to show $I(\mathbf{x};\mathbf{y})=I(\mathbf{x}';\mathbf{y}')$. Due
to the circularity of $\mathbf{H}$, Theorem \ref{a1_charac_div} implies,
\begin{align*}
    h\left(\mathbf{H}\mathbf{x} +\mathbf{z}\right)=h\left(\mathbf{H}e^{\jmath 2 \pi \psi}
\mathbf{x}' +\mathbf{z}\right)=h\left(\mathbf{H}\mathbf{x}' +\mathbf{z}\right),
\end{align*}
so that it remains to show $h\left(\mathbf{H}\mathbf{x}
+\mathbf{z}|\mathbf{x}\right)=h\left(\mathbf{H}\mathbf{x}'
+\mathbf{z}|\mathbf{x}'\right)$. Fubini's Theorem yields
\begin{align*}
  h\left(\mathbf{H}\mathbf{x}+\mathbf{z}|\mathbf{x}\right)&= \int h\left(\mathbf{H}\mathbf{x}+\mathbf{z}\big|\mathbf{x}^{\text{(r)}}=\xxi\right) f_{\mathbf{x}^{\text{(r)}}}(\xxi) d\xxi\\
  &= \int_{0}^{1}\int h\left(\mathbf{H}\mathbf{x}+\mathbf{z}\big|\mathbf{x}^{\text{(r)}}=\xxi\right) f_{(e^{\jmath 2
\pi \varphi}\mathbf{x}')^{\text{(r)}}}(\xxi) d\xxi d\varphi,
\end{align*}
since $f_{\mathbf{x}^{\text{(r)}}}(\xxi)=\int_0^1
f_{\mathbf{x}^{\text{(r)}}|\psi}(\xxi|\varphi) d\varphi = \int_0^1
f_{(e^{\jmath 2 \pi \varphi}\mathbf{x}')^{\text{(r)}}}(\xxi) d\varphi$, and,
furthermore,
\begin{align*}
  h\left(\mathbf{H}\mathbf{x}+\mathbf{z}|\mathbf{x}\right)&= \int_{0}^{1}\int h\left(\mathbf{H}e^{\jmath 2
\pi \varphi}\mathbf{x}'+\mathbf{z}\big|{\mathbf{x}'}^{\text{(r)}}=\xxi\right) f_{\mathbf{x}'^{\text{(r)}}}(\xxi) d\xxi d\varphi\\
&\overset{(*)}{=} \int_{0}^{1}\int h\left(\mathbf{H}\mathbf{x}'+\mathbf{z}\big|{\mathbf{x}'}^{\text{(r)}}=\xxi\right) f_{\mathbf{x}'^{\text{(r)}}}(\xxi) d\xxi d\varphi\\
&= \int_{0}^{1}h\left(\mathbf{H}\mathbf{x}'
+\mathbf{z}|\mathbf{x}'\right) d\varphi\\
&=h\left(\mathbf{H}\mathbf{x}'
+\mathbf{z}|\mathbf{x}'\right),
\end{align*}
where $(*)$ follows from the circularity of $\mathbf{H}$. Hence, the circular
$\mathbf{x}$ achieves the noncoherent capacity of
\eqref{MIMO_channel_model}.\eproof
\begin{theorem}\label{Th:cap5}
  Suppose for \eqref{MIMO_channel_model} a random channel matrix \emph{$\mathbf{H}\in
\C^{n \times m}$}, such that the random vector, which is obtained from an
arbitrary stacking of the elements of $\mathbf{H}$ into an $n m$-dimensional
vector, is circular, and a circular-closed set \emph{$\mathcal{I}$} of
admissible input
  distributions. Then, there exists a circular random
  vector $\mathbf{x}\in\C^{m}$ that achieves the coherent capacity of \eqref{MIMO_channel_model_coherent}.
\end{theorem}\bproof Let us denote by $\mathbf{x}'\in\mathcal{I}\, $ a---not necessarily circular---capacity-achieving random
vector and let $\mathbf{ x}_{(\theta)} \triangleq e^{\jmath 2 \pi
\theta}\mathbf{x}'$ (with $\theta\in [0,1[$ being deterministic). With
$\mathbf{y}_{(\theta)} =\mathbf{H} \mathbf{x}_{(\theta)}  +\mathbf{z}$ we
obtain,
\begin{align*}
  I\left(\mathbf{x}_{(\theta)};\mathbf{y}_{(\theta)},\mathbf{H}\right)&=
  \int I\left(\mathbf{x}^{\text{(r)}}_{(\theta)};\mathbf{y}^{\text{(r)}}_{(\theta)}\big|\mathbf{H}^{\text{(r)}}=\cchi\right) f_{\mathbf{H}^{\text{(r)}}}(\cchi) d\cchi\\
  &=\int \left( h\left(\mathbf{H} \mathbf{x}_{(\theta)}  +\mathbf{z} \big|\mathbf{H}^{\text{(r)}}=\cchi \right) - h\left(\left(\mathbf{H}\mathbf{x}_{(\theta)} +\mathbf{z}\big|\mathbf{x}_{(\theta)}\right)\big|\mathbf{H}^{\text{(r)}}=\cchi\right) \right) f_{\mathbf{H}^{\text{(r)}}}(\cchi) d\cchi\\
  &=\int \left( h\left(\mathbf{H} e^{\jmath 2 \pi
\theta}\mathbf{x}'  +\mathbf{z} \big|\mathbf{H}^{\text{(r)}}=\cchi \right) - h\left(\left(\mathbf{H}e^{\jmath 2 \pi
\theta}\mathbf{x}' +\mathbf{z}\big|\mathbf{x}'\right)\big|\mathbf{H}^{\text{(r)}}=\cchi\right) \right) f_{\mathbf{H}^{\text{(r)}}}(\cchi) d\cchi\\
&=\int \left( h\left(\mathbf{H} \mathbf{x}'  +\mathbf{z} \big|\mathbf{H}^{\text{(r)}}=\cchi \right) - h\left(\left(\mathbf{H}\mathbf{x}' +\mathbf{z}\big|\mathbf{x}'\right)\big|\mathbf{H}^{\text{(r)}}=\cchi\right) \right) f_{(e^{\jmath 2 \pi
\theta}\mathbf{H})^{\text{(r)}}}(\cchi) d\cchi,
\end{align*}
where Fubini's Theorem has been used, and, furthermore, due to the
circularity of $\mathbf{H}$,
\begin{align*}
  I\left(\mathbf{x}_{(\theta)};\mathbf{y}_{(\theta)},\mathbf{H}\right)&=
  \int I\left({\mathbf{x}'}^{\text{(r)}};{\mathbf{y}'}^{\text{(r)}}\big|\mathbf{H}^{\text{(r)}}=\cchi\right) f_{\mathbf{H}^{\text{(r)}}}(\cchi) d\cchi=I\left(\mathbf{x}';\mathbf{y}',\mathbf{H}\right),
\end{align*}
where $\mathbf{y}' =\mathbf{H} \mathbf{x}'  +\mathbf{z}$. Hence, $\mathbf{
x}_{(\theta)}$ is capacity-achieving. Therefore, by applying Jensen's
inequality to the concave mutual information function (with respect to the
input distribution, cf. the proof of Theorem \ref{Th:cap2}), the random
vector $\mathbf{x}\in\C^{m}$ with distribution defined according to
$f_{\mathbf{x}^{\text{(r)}}}(\xxi)\triangleq\int_0^1
f_{\mathbf{x}_{(\theta)}^{\text{(r)}}}(\xxi)
d\theta$\hspace*{1mm}$\lambda_{2n}\text{-a.e.}$ satisfies
\begin{align*}
  I\left(\mathbf{x};\mathbf{y},\mathbf{H}\right) \geq \int_0^1 I\left(\mathbf{x}_{(\theta)};\mathbf{y}_{(\theta)},\mathbf{H}\right) d\theta =
  \int_0^1 I\left(\mathbf{x}';\mathbf{y}',\mathbf{H}\right) d\theta = I\left(\mathbf{x}';\mathbf{y}',\mathbf{H}\right),
\end{align*}
so that $\mathbf{x}$ achieves the coherent capacity of
\eqref{MIMO_channel_model_coherent}. But
$f_{\mathbf{x}_{(\theta)}^{\text{(r)}}}(\xxi)
=f_{{\mathbf{x}'}_{\text{(a)}}^{\text{(r)}}|\psi}(\xxi|\theta)$\hspace*{1mm}$\lambda_{2n}\text{-a.e.}$,
where $\psi$ denotes the uniformly distributed random variable used for
defining ${\mathbf{x}'}_{\text{(a)}}$ (see Definition \ref{def:a1}), and,
therefore, $\mathbf{x}={\mathbf{x}'}_{\text{(a)}} \in \mathcal{I}$. \eproof

\subsection{Deterministic Channel Matrix and Improper Gaussian Noise Vector}\label{sec:improp_noise}
Here, we investigate the case that the channel matrix $\mathbf{H} \in \C^{n
\times m}$ is deterministic and that the noise vector $\mathbf{z}\in\C^{n}$
is Gaussian distributed. We impose an average power constraint, i.e., we
define the set of admissible input distributions as
\begin{align}\label{power_constraint}
  \mathcal{I}\triangleq\{\mathbf{x}: \E{\mathbf{x}^{H}\mathbf{x}} \leq S\}.
\end{align}
For $\mathbf{z}$ proper, both capacity and capacity-achieving input vector
are well known \cite{Telatar95}. Therefore, in the following, we consider a
more general situation without the assumption of $\mathbf{z}$ being proper.
However, we introduce additional technical assumptions, which make the
derivation less complicated and lead to simpler results. Note, that most of
these assumptions could be significantly relaxed or even omitted, but for the
price of more involved theorems and proofs. We assume,
\begin{subequations}\label{tech_ass_all}
\begin{eqnarray}\label{tech_ass_1}
  &\mathbf{H} \in \C^{n \times n} \hspace*{2mm} \text{deterministic, quadratic, and non-singular},&\\\label{tech_ass_2}
  &S \geq 2 n \left\|\mathbf{H}^{-1} \mathbf{C}_{\mathbf{z}} \mathbf{H}^{-H}\right\|_2 \hspace*{2mm} \text{(high signal-to-noise ratio)},&\\\label{tech_ass_3}
  &\mathbf{z} \hspace*{2mm} \text{zero-mean with non-singular} \hspace*{2mm} \mathbf{C}_{\mathbf{z}} \in \C^{n \times n},&\\\label{tech_ass_4}
  &\left\|\mathbf{B}_{\mathbf{z}}^{-1}\mathbf{P}_{\mathbf{z}}\mathbf{B}_{\mathbf{z}}^{-T}\right\|_{2}<1,&
\end{eqnarray}
\end{subequations}
where $\mathbf{C}_{\mathbf{z}}$ and $\mathbf{P}_{\mathbf{z}}$ denote
covariance matrix and complementary covariance matrix of $\mathbf{z}$,
respectively, and $\mathbf{B}_{\mathbf{z}}$ is a generalized Cholesky factor
of $\mathbf{C}_{\mathbf{z}}$. We have the following capacity result.
\begin{theorem}\label{Th:cap6}
Suppose for \eqref{MIMO_channel_model} that assumptions
\eqref{tech_ass_all} hold and that the set of admissible input distributions
is defined according to \eqref{power_constraint}. Then, the capacity of
\eqref{MIMO_channel_model} is given by\emph{
\begin{align*}
  C&= 2\log \left|\det \mathbf{H}\right|+n\log\left(S + \tr{\mathbf{H}^{-1} \mathbf{C}_{\mathbf{z}} \mathbf{H}^{-H}} \right) - \log \det \mathbf{C}_{\mathbf{z}} -
  \frac{1}{2} \sum\limits_{i=1}^{n} \log(1-\lambda_{i}^{2})- n \log n,
\end{align*}}
where \emph{$\tr{\cdot}$} denotes the usual matrix trace and \emph{$\lambda_{i}$}
are the singular values of
\emph{$\mathbf{B}_{\mathbf{z}}^{-1}\mathbf{P}_{\mathbf{z}}\mathbf{B}_{\mathbf{z}}^{-T}$}.
Furthermore, the zero-mean and Gaussian distributed random vector
\emph{$\mathbf{x} \in \C^{n}$} with covariance matrix and complementary
covariance matrix given by\emph{
\begin{subequations}\label{cov_compmat_max}
\begin{align}\label{covmat_max}
\mathbf{C}_{\mathbf{x}}&=
\frac{1}{n}\left(S + \tr{\mathbf{H}^{-1}
\mathbf{C}_{\mathbf{z}} \mathbf{H}^{-H}} \right) \mathbf{I}_n -\mathbf{H}^{-1}
\mathbf{C}_{\mathbf{z}} \mathbf{H}^{-H},\\\label{compmat_max}
\mathbf{P}_{\mathbf{x}}&= -\mathbf{H}^{-1}
\mathbf{P}_{\mathbf{z}} \mathbf{H}^{-T},
\end{align}
\end{subequations}} respectively, is capacity-achieving.
\end{theorem}
\bproof Since $\E{\mathbf{x}^{H}\mathbf{x}} = \tr{\mathbf{C}_{\mathbf{x}}}+
\|\mathbf{m}_{\mathbf{x}}\|^{2}_{2}$, where $\mathbf{m}_{\mathbf{x}}$ denotes
the mean vector of $\mathbf{x}$, Theorem \ref{max_entropy_th_neeser} implies
that the supremum of
\begin{align*}
  I(\mathbf{x};\mathbf{y}) &= h(\mathbf{y}) - h\left(\mathbf{y}|\mathbf{x}\right)= h\left(\mathbf{H}\mathbf{x} +\mathbf{z}\right) - h\left(\mathbf{z}\right)
\end{align*}
over $\mathcal{I}$ is achieved by a zero-mean and Gaussian distributed
complex-valued random vector $\mathbf{x}$ with covariance matrix
$\mathbf{C}_{\mathbf{x}}$ that maximizes the function $g\left(\mathbf{C}
\right)\triangleq\log \det \left(\mathbf{H}\mathbf{C}\mathbf{H}^{H}
+\mathbf{C}_{\mathbf{z}} \right)$ over the set of covariance matrices
$\mathbf{C}$ with $\tr{\mathbf{C}}\leq S$, and with complementary covariance
matrix $\mathbf{P}_{\mathbf{x}}$ that satisfies
$\mathbf{H}\mathbf{P}_{\mathbf{x}}\mathbf{H}^{T} +\mathbf{P}_{\mathbf{z}} =
\mathbf{0}$, provided that such a random vector exists. Using the eigenvalue
decomposition $\mathbf{H}^{-1} \mathbf{C}_{\mathbf{z}} \mathbf{H}^{-H} =
\mathbf{U} \mathbf{D} \mathbf{U}^{H}$ we obtain,
\begin{align*}
    g\left(\mathbf{C}\right)&=2\log \left|\det \mathbf{H}\right|+\log\det\left(\mathbf{C}
+\mathbf{H}^{-1}\mathbf{C}_{\mathbf{z}}\mathbf{H}^{-H} \right)\\
&=2\log \left|\det \mathbf{H}\right|+\log\det\left(\mathbf{C}
+\mathbf{U} \mathbf{D} \mathbf{U}^{H} \right)\\
&=2\log \left|\det \mathbf{H}\right|+\log\det\left(\mathbf{U}^{H}\mathbf{C}\mathbf{U}
+ \mathbf{D}  \right),
\end{align*}
so that its maximum is achieved at $\mathbf{C} =
\mathbf{C}_{\mathbf{x}}\triangleq \mathbf{U} \left(L
\mathbf{I}_{n}-\mathbf{D}\right) \mathbf{U}^{H} = L \mathbf{I}_{n} -
\mathbf{H}^{-1} \mathbf{C}_{\mathbf{z}} \mathbf{H}^{-H}$, where $L$ is chosen
such that
\begin{align}\label{water_level}
  S&=\tr{\mathbf{C}_{\mathbf{x}}}= L n - \tr{\mathbf{H}^{-1}\mathbf{C}_{\mathbf{z}}\mathbf{H}^{-H}}
\end{align}
is satisfied. Note that this is the well-known \emph{water filling} solution
\cite{Telatar95,Cover91}, with the additional simplification that
$\mathbf{C}_{\mathbf{x}}$ is non-singular,\footnote{The \emph{water level}
$L$ is larger than the noise power for all (parallel) \emph{eigenchannels}.}
since, according to \eqref{tech_ass_2},
\begin{align}\label{power_inequ}
  \frac{S}{n}\geq 2  \left\|\mathbf{H}^{-1} \mathbf{C}_{\mathbf{z}} \mathbf{H}^{-H}\right\|_2= 2  \left\|\mathbf{D}\right\|_2 = 2 d_{\max} > d_{\max},
\end{align}
where $d_{\max}$ is the largest entry (eigenvalue) of $\mathbf{D}$. This
yields \eqref{covmat_max}. Clearly, the choice \eqref{compmat_max} satisfies
$\mathbf{H}\mathbf{P}_{\mathbf{x}}\mathbf{H}^{T} +\mathbf{P}_{\mathbf{z}} =
\mathbf{0}$. It remains to show that
$\{\mathbf{C}_{\mathbf{x}},\mathbf{P}_{\mathbf{x}}\}$ is a valid pair of
covariance matrix and complementary covariance matrix. To that end, consider
\begin{subequations}
\begin{align}\label{Px_bound1}
  \left\|\mathbf{P}_{\mathbf{x}} \right\|_2 &= \left\|\mathbf{H}^{-1}
\mathbf{P}_{\mathbf{z}} \mathbf{H}^{-T}\right\|_2\\\nonumber
&= \left\|\mathbf{H}^{-1} \mathbf{B}_{\mathbf{z}}\left( \mathbf{B}_{\mathbf{z}}^{-1}
\mathbf{P}_{\mathbf{z}}\mathbf{B}_{\mathbf{z}}^{-T}\right)\mathbf{B}_{\mathbf{z}}^{T}\mathbf{H}^{-T}\right\|_2\\\nonumber
&\overset{(*)}{\leq} \left\|\mathbf{H}^{-1} \mathbf{B}_{\mathbf{z}}\right\|_{2}^{2}\\\nonumber
&=\left\|\mathbf{H}^{-1} \mathbf{C}_{\mathbf{z}} \mathbf{H}^{-H}\right\|_2\\\nonumber
&=   \left\|\mathbf{D}\right\|_2\\\label{Px_bound2}
&= d_{\max},
\end{align}
\end{subequations} where $(*)$ follows from Theorem \ref{pseud_criterion}, and
note that $L>\frac{S}{n}\geq 2 d_{\max}$, cf. \eqref{water_level} and
\eqref{power_inequ}. This implies,
\begin{align*}
\left\|\mathbf{P}_{\mathbf{x}} \right\|_2 < L- d_{\max} = \frac{1}{\left\|\mathbf{C}_{\mathbf{x}}^{-1}\right\|_{2}},
\end{align*}
and, furthermore,
\begin{align*}
\left\|\mathbf{B}_{\mathbf{x}}^{-1}\mathbf{P}_{\mathbf{x}} \mathbf{B}_\mathbf{x}^{-T}\right\|_2
\leq \left\|\mathbf{C}_{\mathbf{x}}^{-1}\right\|_2 \left\|\mathbf{P}_{\mathbf{x}} \right\|_2 < 1,
\end{align*}
so that Theorem \ref{pseud_criterion} shows that \eqref{cov_compmat_max}
defines a valid pair of covariance matrix and complementary covariance
matrix. Finally, the capacity of \eqref{MIMO_channel_model} is obtained as
\begin{align*}
  C&=g\left(\mathbf{C}_{\mathbf{x}}\right) + n \log (\pi e) - h\left(\mathbf{z}\right)\\
  &=2\log \left|\det \mathbf{H}\right|+n \log \left(
\frac{1}{n}\left(S + \tr{\mathbf{H}^{-1}
\mathbf{C}_{\mathbf{z}} \mathbf{H}^{-H}} \right)\right) + n \log (\pi e) - \log \det (\pi e \mathbf{C}_{\mathbf{z}})-\frac{1}{2} \sum\limits_{i=1}^{n} \log(1-\lambda_{i}^{2})\\
  &=2\log \left|\det \mathbf{H}\right|+n \log
\left(S + \tr{\mathbf{H}^{-1}
\mathbf{C}_{\mathbf{z}} \mathbf{H}^{-H}} \right) - n \log n - \log \det \mathbf{C}_{\mathbf{z}}-\frac{1}{2} \sum\limits_{i=1}^{n} \log(1-\lambda_{i}^{2}),
\end{align*}
where Theorem \ref{max_entropy_th_taubII} has been used.\eproof {\bf
Remarks.} Whereas in many real-world scenarios the noise vector happens to be
circular, so that the results of Section \ref{sec:circ_noise} apply, there
are also practically relevant scenarios, where the noise vector is known to
be improper. More specifically, DMT modulation, which is widely used in xDSL
applications \cite{xDSL_overview97}, yields an equivalent system channel that
exactly matches the situation considered here
\cite{taub_comp_glob03,taub_comp_sp07}. We also note that capacity results
for improper Gaussian distributed noise vectors could be alternatively
derived by making use of an equivalent real-valued channel of dimension $2n
\times 2m$ that is obtained by appropriate stacking of real and imaginary
parts. The advantage of the approach presented here is that it yields
expressions that are explicit in covariance matrix and complementary
covariance matrix. In the following, we make use of this (desired)
separation.

Observe that improper noise is beneficial since---due to $\frac{1}{2}
\sum\limits_{i=1}^{n} \log(1-\lambda_{i}^{2}) < 0$---it increases capacity.
However, this presupposes a suitably designed transmission scheme. If it is
erroneously believed that $\mathbf{P}_{\mathbf{z}}= \mathbf{0}$, it will be
erroneously believed as well (see Theorem \ref{Th:cap6}) that the zero-mean
and Gaussian distributed random vector $\mathbf{x}'$ with covariance matrix
$\mathbf{C}_{\mathbf{x}'}=\mathbf{C}_{\mathbf{x}}$ as in \eqref{covmat_max}
but with $\mathbf{P}_{\mathbf{x}'}= \mathbf{0}$ is capacity-achieving. It
follows that
\begin{align*}
    C' \triangleq I(\mathbf{x}';\mathbf{y}') = C - \Delta C \leq C,
\end{align*}
where $\mathbf{y}'=\mathbf{H} \mathbf{x}' +\mathbf{z}$ and $\Delta C
\triangleq C-C'$ denotes the resulting \emph{capacity loss}. This capacity
loss is quantified by the next theorem.
\begin{theorem}\label{Th:caploss}
Suppose for \eqref{MIMO_channel_model} that assumptions \eqref{tech_ass_all}
hold and that the set of admissible input distributions is defined according
to \eqref{power_constraint}. Then, the capacity loss $\Delta C$ that occurs
if it is erroneously believed that $\mathbf{P}_{\mathbf{z}}= \mathbf{0}$ is
given by \emph{
\begin{align*}
  \Delta C = - \frac{1}{2} \sum\limits_{i=1}^{n} \log(1-\mu_{i}^{2}),
\end{align*}}
where \emph{$\mu_{i}$} are the singular values of \emph{$\left(n/\left(S +
\tr{\mathbf{H}^{-1} \mathbf{C}_{\mathbf{z}}
\mathbf{H}^{-H}}\right)\right)\mathbf{H}^{-1}\mathbf{P}_{\mathbf{z}}\mathbf{H}^{-T}$}.
In particular, \emph{
\begin{align*}
  0 \leq \Delta C  < n \log \frac{2}{\sqrt{3}}.
\end{align*}}
\end{theorem}
\bproof We intend to apply Theorem \ref{max_entropy_th_taubII} to the random
vector $\mathbf{y}'=\mathbf{H} \mathbf{x}' +\mathbf{z}$. In order to meet the
assumption of Theorem \ref{max_entropy_th_taubII}, we have to show
$\|\mathbf{B}_{\mathbf{y}'}^{-1}\mathbf{P}_{\mathbf{y}'}\mathbf{B}_{\mathbf{y}'}^{-T}\|_{2}<1$,
where $\mathbf{B}_{\mathbf{y}'}$ denotes a generalized Cholesky factor of
$\mathbf{C}_{\mathbf{y}'}$. Note that the non-singularity of
$\mathbf{C}_{\mathbf{y}'}$ follows from the non-singularity of
$\mathbf{C}_{\mathbf{z}}$. Clearly, $\mathbf{C}_{\mathbf{y}'}=\mathbf{H}
\mathbf{C}_{\mathbf{x}} \mathbf{H}^{H} + \mathbf{C}_{\mathbf{z}}=
\frac{1}{n}\left(S + \tr{\mathbf{H}^{-1} \mathbf{C}_{\mathbf{z}}
\mathbf{H}^{-H}} \right) \mathbf{H} \mathbf{H}^{H}$ and
$\mathbf{P}_{\mathbf{y}'}= \mathbf{P}_{\mathbf{z}}$, so that
$\mathbf{B}_{\mathbf{y}'}^{-1}\mathbf{P}_{\mathbf{y}'}\mathbf{B}_{\mathbf{y}'}^{-T}=
\left(n/\left(S + \tr{\mathbf{H}^{-1} \mathbf{C}_{\mathbf{z}}
\mathbf{H}^{-H}}\right)\right)\mathbf{H}^{-1}\mathbf{P}_{\mathbf{z}}\mathbf{H}^{-T}$
and, furthermore,
\begin{eqnarray*}
  \|\mathbf{B}_{\mathbf{y}'}^{-1}\mathbf{P}_{\mathbf{y}'}\mathbf{B}_{\mathbf{y}'}^{-T}\|_{2}\hspace*{-3mm} &=& \hspace*{-3mm}
    \frac{n}{S + \tr{\mathbf{H}^{-1}
\mathbf{C}_{\mathbf{z}} \mathbf{H}^{-H}} } \|\mathbf{H}^{-1}\mathbf{P}_{\mathbf{z}}\mathbf{H}^{-T}\|_{2}\\
\hspace*{-3mm}&\overset{\eqref{Px_bound1},\eqref{Px_bound2}}{\leq}&\hspace*{-3mm}\frac{n}{S + \tr{\mathbf{H}^{-1}
\mathbf{C}_{\mathbf{z}} \mathbf{H}^{-H}} } \,d_{\max}\\
\hspace*{-3mm}&<&\hspace*{-3mm}\frac{n}{S} \,\,d_{\max}\\
\hspace*{-3mm}&\overset{\eqref{power_inequ}}{\leq}&\hspace*{-3mm} \frac{1}{2}.
\end{eqnarray*}
The capacity loss is then given by
\begin{align*}
  \Delta C &= C-C'\\
  &= h\left(\mathbf{y}\right) - h\left(\mathbf{z}\right) - h\left(\mathbf{y}'\right) + h\left(\mathbf{z}\right)\\
  &= h\left(\mathbf{y}\right) -  h\left(\mathbf{y}'\right)\\
  &\overset{(*)}{=} \log \det (\pi e \mathbf{C}_{\mathbf{y}})
  -\log \det (\pi e \mathbf{C}_{\mathbf{y}})-\frac{1}{2} \sum\limits_{i=1}^{n} \log(1-\mu_{i}^{2})\\
  &=-\frac{1}{2} \sum\limits_{i=1}^{n} \log(1-\mu_{i}^{2}),
\end{align*}
where $(*)$ follows from Theorem \ref{max_entropy_th_taubII}, since
$\mathbf{P}_{\mathbf{y}} = \mathbf{0}$ and
$\mathbf{C}_{\mathbf{y}'}=\mathbf{C}_{\mathbf{y}}$. For the bound note that
\begin{align*}
  -\frac{1}{2} \sum\limits_{i=1}^{n} \log(1-\mu_{i}^{2}) &\leq -\frac{n}{2} \log(1-\|\mathbf{B}_{\mathbf{y}'}^{-1}\mathbf{P}_{\mathbf{y}'}\mathbf{B}_{\mathbf{y}'}^{-T}\|_{2}^{2}) <
    -\frac{n}{2}  \log(1-\frac{1}{4}) = n \log \frac{2}{\sqrt{3}}.
\end{align*}
\eproof \unitlength1mm
\begin{figure}
\begin{center}
\begin{picture}(85,98)(0,-13)
\put(-10,9){\includegraphics{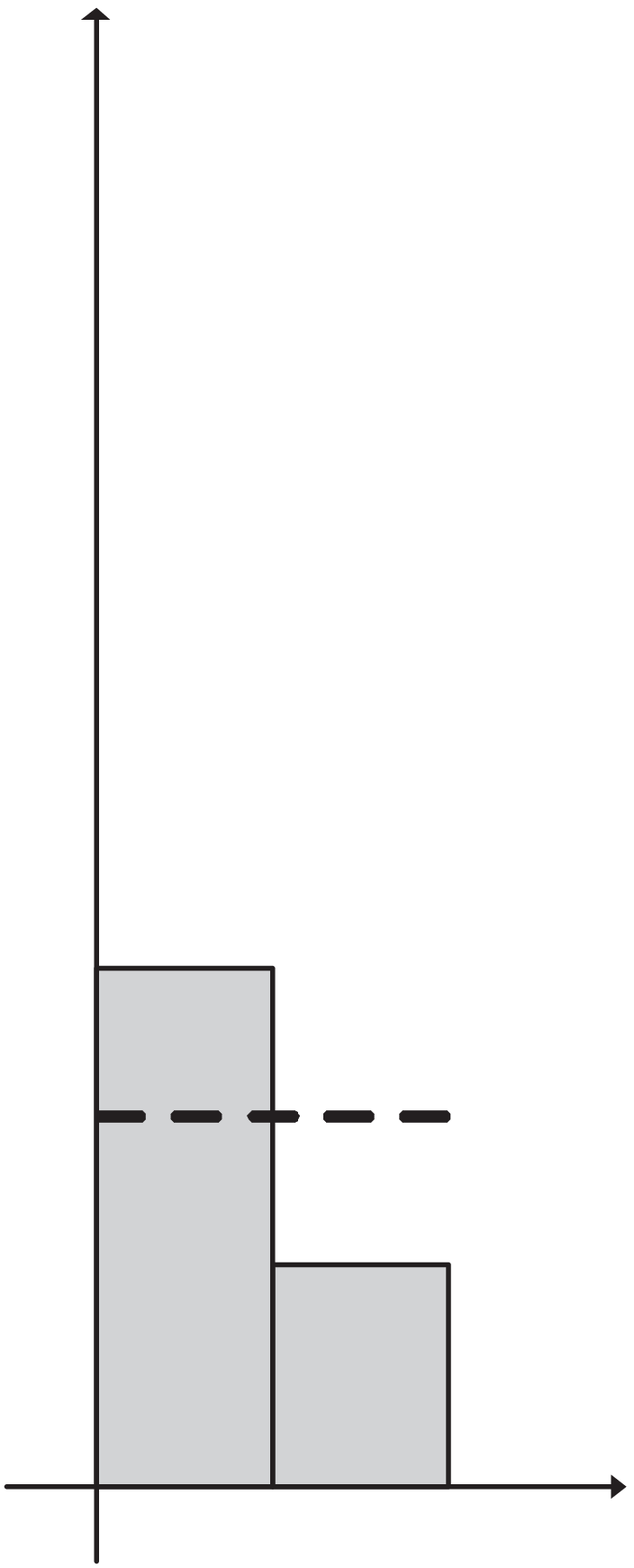} } \put(30,9){\includegraphics{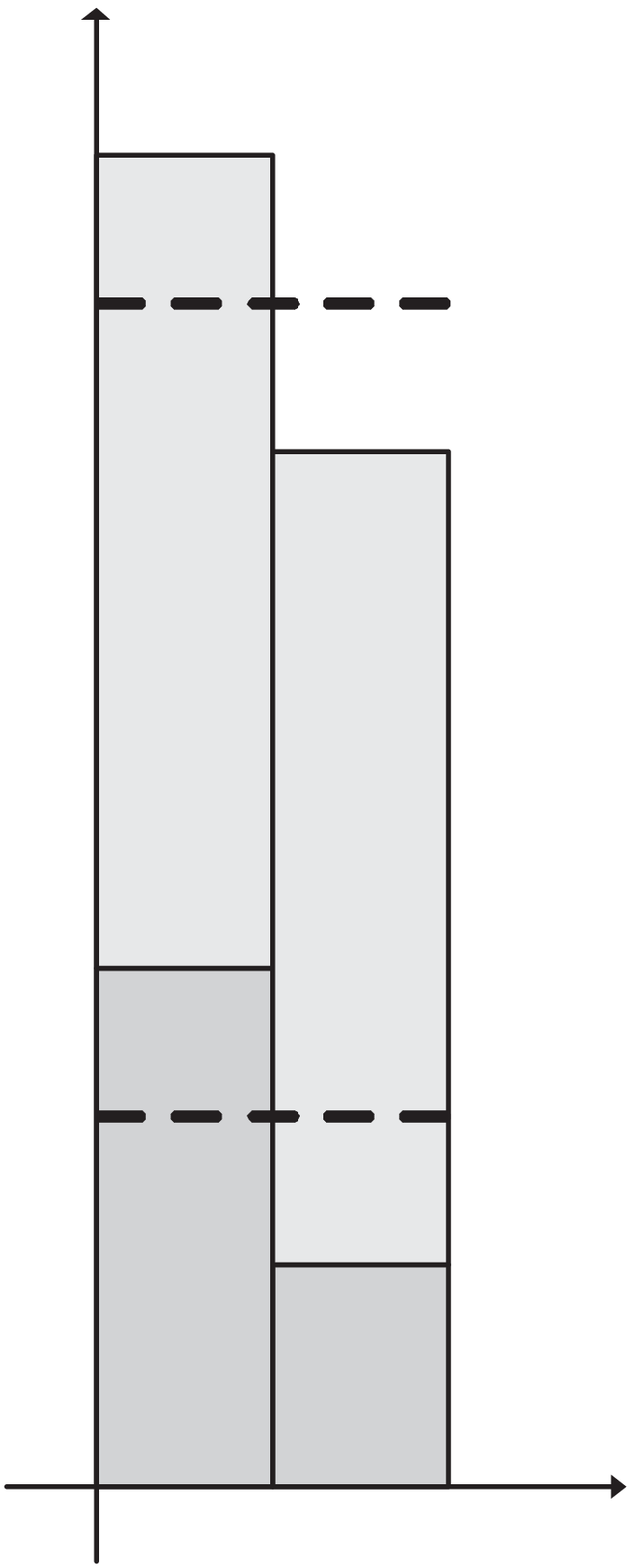} }
\put(70,9){\includegraphics{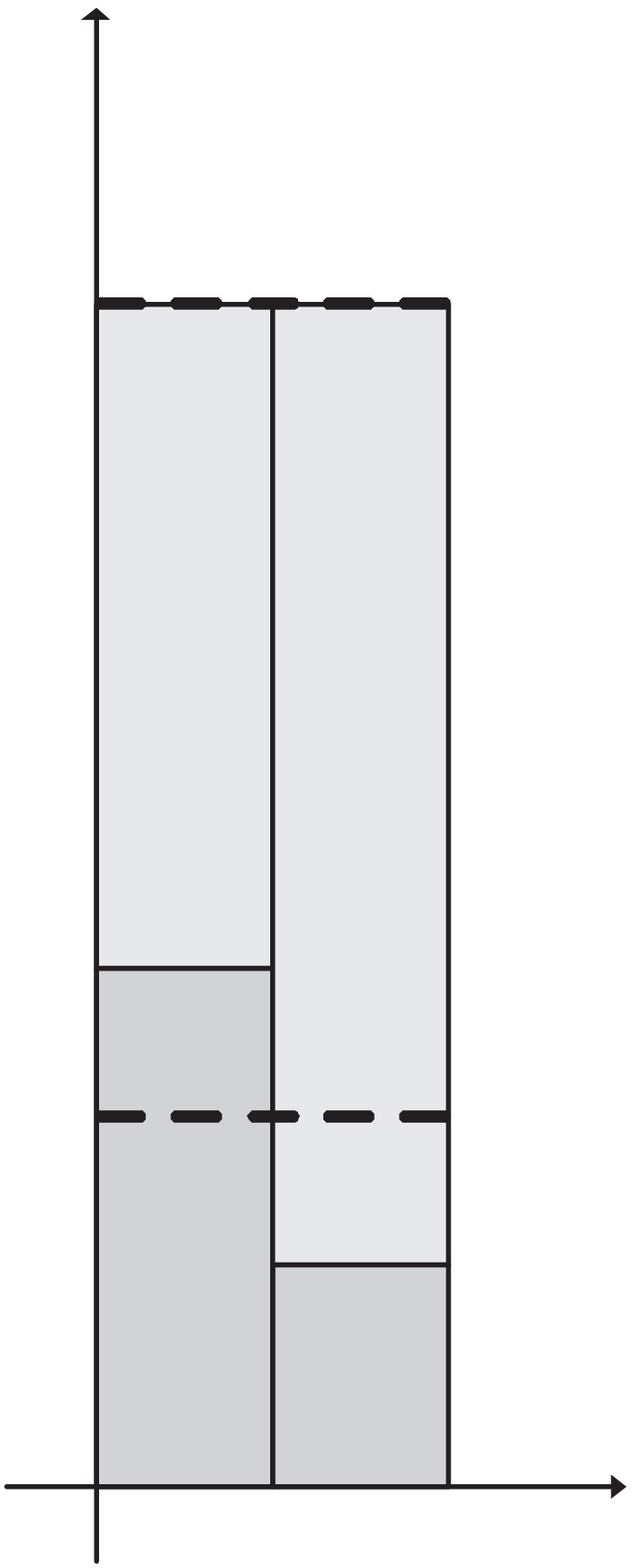} } \put(-0.5,1){{(a)}} \put(39.5,1){{(b)}}
\put(79.5,1){{(c)}} \put(-3,8){{$\RE$}} \put(5,8){{$\IM$}}
\put(37,8){{$\RE$}} \put(45,8){{$\IM$}} \put(77,8){{$\RE$}}
\put(85,8){{$\IM$}}

\put(-17,76){{\small power}} \put(23,76){{\small power}} \put(63,76){{\small
power}}

\put(-13,28.3){{\small $\frac{C_{z}}{2}$}} \put(-16,35.2){{\small
$\frac{C_{z}+P_{z}}{2}$}} \put(-16,21.6){{\small $\frac{C_{z}-P_{z}}{2}$}}

\put(27,28.3){{\small $\frac{C_{z}}{2}$}} \put(24,35.2){{\small
$\frac{C_{z}+P_{z}}{2}$}} \put(24,21.6){{\small $\frac{C_{z}-P_{z}}{2}$}}
\put(24,65.1){{\small $\frac{S+C_{z}}{2}$}}

\put(67,28.3){{\small $\frac{C_{z}}{2}$}} \put(64,35.2){{\small
$\frac{C_{z}+P_{z}}{2}$}} \put(64,21.6){{\small $\frac{C_{z}-P_{z}}{2}$}}
\put(64,65.1){{\small $\frac{S+C_{z}}{2}$}}

\put(-7,-5){{\small noise power}} \put(-7,-10){{\small distribution}}
\put(35,-5){{\small erroneous}} \put(33.5,-10){{\small water
filling}}\put(77,-5){{\small {correct }}} \put(73.5,-10){{\small water
filling}}

\end{picture}
\caption{Water filling strategies illustrating the capacity loss.}
\label{Fig_CapLoss}
\end{center}
\end{figure}
{\bf Example.} Let us consider the special case of the complex scalar channel
$y=x+z$ with noise covariance $C_{z} \in \R$ and complementary noise
covariance $P_{z} \in \R$, where $C_{z} \geq P_{z} > 0$. According to
\eqref{expr_cov_compl},
\begin{align*}
\mathbf{C}_{z^{\text{(r)}}} = \frac{1}{2}\left[
\begin{array}{cc}
  C_{z}+P_{z} & 0 \\
  0 & C_{z}-P_{z} \\
\end{array}%
\right],
\end{align*}
which is illustrated in Fig.~\ref{Fig_CapLoss}(a). It is seen that the noise
power is different for real and imaginary part. If it is erroneously believed
that $P_{z} = 0$, the same power is assigned to real and imaginary part of
the input vector, as it is shown in Fig.~\ref{Fig_CapLoss}(b). However, the
optimum power distribution that maximizes the mutual information is
different; it is depicted in Fig.~\ref{Fig_CapLoss}(c). Note that this
capacity-achieving power distribution is obtained by water filling on a real
and imaginary part level. The difference between the mutual informations of
solution (b) and (c) is expressed by the capacity loss $\Delta C$.

\section{Conclusion}\label{sec:conc}

We studied the influence of circularity/non-circularity and
propeness/improperness on important information theoretic quantities such as
entropy, divergence, and capacity. As a motivating starting point served a
theorem by \emph{Neeser} \& \emph{Massey} \cite{neeser93}, which states that
the entropy of a zero-mean complex-valued random vector is upper-bounded by
the entropy of a circular/proper Gaussian distributed random vector with same
covariance matrix. We strengthened this theorem in two different directions:
(i) we dropped the Gaussian assumption and (ii) we dropped the properness
assumption. In both cases the resulting upper-bound turned out to be tighter
than the one previously known. A key ingredient for the proof in case (i) was
the introduction of the \emph{circular analog} of a given complex-valued
random vector. Whereas its definition was based on intuitive arguments to
obtain a circular random vector, which is ``close'' to the (potentially)
non-circular given one, we rigorously proved that it equals the unique
circular random vector with minimum Kullback-Leibler divergence. On the other
hand, for (ii), we exploited results about the second-order structure of
complex-valued random vectors that were obtained without making use of the
augmented covariance matrix (in contrast to related work). Additionally, we
presented a criterion for a matrix to be a valid complementary covariance
matrix. Furthermore, we addressed the capacity of MIMO channels. Regardless
of the specific distribution of the channel parameters (noise vector and
channel matrix, if modeled as random), we showed that the capacity-achieving
input vector is circular for a broad range of MIMO channels (including
coherent and noncoherent scenarios). This extends known results that make use
of a Gaussian assumption. Finally, we investigated the situation of an
improper and Gaussian distributed noise vector. We computed both capacity and
capacity-achieving input vector and showed that improperness increases
capacity, provided that the complementary covariance matrix is exploited.
Otherwise, a capacity loss occurs, for which we derived an explicit
expression.

\section*{Acknowledgments}\label{sec.ack}

The author would like to thank J.\ Huber, J.\ Sayir, and J.\ Weinrichter for
helpful hints and comments. He is also grateful to the anonymous reviewers
for their constructive comments that have resulted in a major improvement of
this paper.


\renewcommand{\baselinestretch}{1.09}\small\normalsize\small
\bibliographystyle{ieeetr}

\end{document}